\documentclass[12pt,preprint]{aastex}

\newcommand{\gapprox}{\hbox{\lower .8ex\hbox{$\,\buildrel > \over\sim\,$}}}
\newcommand{\lapprox}{\hbox{\lower .8ex\hbox{$\,\buildrel < \over\sim\,$}}}
\newcommand\mathhyphen{\hbox{-}}
\newcount\countitemc
\def\itemcnumber{\number\countitemc}
\def\itemc{\global\advance\countitemc by 1   \itemcnumber}

\shorttitle{ON SN~2003fg} 
\shortauthors{Jeffery \&~Branch}

\begin{document}


\title{ON SN~2003fg:  THE PROBABLE SUPER-CHANDRASEKHAR-MASS SN~Ia}

\author{
David J.~Jeffery\altaffilmark{1},
David Branch\altaffilmark{1},
\&
E. Baron\altaffilmark{1}
}



\altaffiltext{1}{Homer L. Dodge Department of Physics \& Astronomy, University of Oklahoma,
                440 W. Brooks St., Norman, Oklahoma 73019, U.S.A.}

\begin{abstract}

        \citeauthor{howell2006} have reported the discovery of SN~Ia SN~2003fg (SNLS-03D3bb) and 
conclude that SN~2003fg is very likely a super-Chandrasekhar-mass SN~Ia perhaps with 
a mass $\sim 2\,M_{\odot}$.    
Their work is the first strong evidence that has been presented for a super-Chandrasekhar SN~Ia.
We have performed an analysis of the SN~2003fg data using the \citeauthor{yoon2005}
binding energy formula for a rotating super-Chandrasekhar-mass white dwarf
(also used by \citeauthor{howell2006}) included in a simple model of SNe~Ia (which we call the
SSC~model for Simple Super-Chandrasekhar model for SNe~Ia) 
which assumes spherically symmetric ejecta and relies 
on the approximations of an exponential density profile
for SN~Ia ejecta and of a sharp boundary of the SN~Ia iron-peak-element core.
Our results support the conclusion of \citeauthor{howell2006}:  SN~2003fg is very
probably super-Chandrasekhar and probably has mass $\sim 2\,M_{\odot}$. 

\end{abstract}



\section{INTRODUCTION\label{section-introduction}}

    Recently, \citet[hereafter H2006]{howell2006},
as part of the Supernova Legacy Survey (SNLS), have reported the discovery of SN~Ia SN~2003fg
(supernova SNLS-03D3bb in the SNLS naming scheme).
SN~2003fg is a remote supernova at $z=0.2440\pm0.0003$ and using concordance cosmology
($H_{0}=70\,{\rm km\,s^{-1}\,Mpc^{-1}}$, $\Omega_{M}=0.3$, flat universe)
is found to be about 2.2 times brighter in absolute $V$ magnitude than the median brightness of SNe~Ia:  it
is intrinsically the brightest SN~Ia known with any confidence.
SN~2003fg's lightcurves in shape are not out of the range of normal SN~Ia behavior. 
Thus, SN~2003fg strongly violates the lightcurve width-luminosity relationship
\citep{phillips1993,phillips1999} which is so useful in cosmology.
 
    The one spectrum published by H2006, which comes from 2 days after $B$ maximum (light)
(which H2006 assume to also be bolometric maximum light),
looks typical of near-$B$-maximum SN~Ia spectra as H2006's Figure~3 shows by a comparison to
the very normal SN~Ia SN~1994D (a core normal SN~Ia in the terminology of \citet{branch2006}).
The spectrum line features, however, are distinctly narrow.
The narrowness suggests somewhat low ejecta velocities for the region of line formation in comparison
to normal SNe~Ia at the same phase.
(The width of the lines is determined by the Doppler shift of line opacity in the
expanding ejecta.)
The line velocity of SN~2003fg's Si~II~$\lambda6355$ line \citep[e.g.,][p.~79]{wiese1969}
(the line which is the most characteristic line of SNe~Ia)
in the spectrum is $8000\pm500\,{\rm km\,s^{-1}}$.
(Line velocity is the Doppler shift velocity corresponding to the wavelength shift of a P~Cygni line
absorption trough minimum from the line center wavelength.
As supernovae evolve the line velocities generally decrease because the ejecta density is falling
and the region of sufficient opacity to form lines recedes into the ejecta in mass fraction and velocity.) 
Core normal SNe~Ia have photospheric velocities
of $\sim 11000\,{\rm km\,s^{-1}}$ at 2 days after $B$ maximum \citep{branch2005,branch2006}.
Since the photospheric velocity is a lower limit on line velocities with normal line formation,
SN~2003fg's photospheric velocity was unusually low for its phase and probably $\sim 8000\,{\rm km\,s^{-1}}$
which is, in fact, the photospheric velocity given by H2006 in their 
supplementary information\footnote{See URL http://www.nature.com/nature/journal/v443/n7109/suppinfo/nature05103.html\ .}

    Note that supernovae after very early times have all mass elements approximately in uniform motion,
and thus all ejecta structures just scale up linearly with time since explosion $t$.
This kind of expansion is called homologous expansion.
In homologous expansion, the radii $r$ of all mass elements obey
\begin{equation}
         r\approx vt  \,\, ,
\label{eq-homologous}
\end{equation}
where $v$ is the mass element velocity and initial radial position in the progenitor is considered negligible.
Because of equation~(\ref{eq-homologous}), velocity becomes a good comoving frame coordinate for
describing supernova ejecta and we conventionally use it as such.
Also in homologous expansion, all mass element densities scale as $t^{-3}$.

    The photospheric velocity is the characteristic velocity of the layer of continuum emission.
In the case of SNe~Ia, the continuum is not a pure blackbody continuum.
Because of falling density and thus opacity, the photosphere, like the line formation region, recedes into
the ejecta in velocity coordinate with passing time.
As the photosphere recedes, the P~Cygni lines gradually transform to emission lines.
Eventually, the photosphere vanishes altogether.
The photospheric phase of a supernova is when the photosphere is still important to the radiative transfer.
After the photospheric phase is the nebular phase.
The transition between the two phases is, of course, gradual.

      One point to emphasize is that the SN~2003fg spectrum in the optical 
was dominated by intermediate-mass element (IME) lines and not iron-peak element (IPE) lines.
(We consider carbon through calcium as IMEs and scandium through nickel as IPEs.)
Optical spectrum dominance in the early post-$B$-maximum phase by IME lines (principally those of Si~II, S~II,
and Ca~II) is true for all SNe~Ia, 
not just normal ones.
This shows that the spectrum formation is not primarily in what we call the IPE core of the ejecta, but 
at higher velocities.
Most SNe~Ia show IME-line dominance in the optical spectra in the pre-$B$-maximum and $B$-maximum phases as well.

      The IPE core is the interior region of many SN~Ia explosion models that is almost entirely IPEs.
At time zero after the explosion the IPE core is in fact mainly radioactive $^{56}$Ni.
The IPE core is a standard feature of explosion models that result from the standard SN~Ia model.
In the standard SN~Ia model, the progenitor is a carbon-oxygen white dwarf (a CO WD with about equal amounts of C and O) 
that is very close to the physical Chandrasekhar mass ($1.38\,M_{\odot}$ for CO WDs with equal amounts of C and O:
see Appendix~\ref{ap-chan-mass}) with typical central densities 
of order $3\times10^{9}\,{\rm g\,cm^{3}}$ and typical central temperatures of order $2.5\times10^{8}\,$K
\citep[e.g.,][p.~108]{woosley1994}.
With such densities and temperatures unstable carbon burning (i.e., carbon ignition) at or near the WD center 
will initiate an explosion that totally disrupts the WD and creates a SN~Ia.

       The WD is driven to the explosive state by mass accretion from a binary companion.
The nature of the companion is not certain at present.
Two possibilities exist (which are not exclusive) for the companion: 
it could be a post-main-sequence star (or very late phase main-sequence star) of some kind  
or it could be another CO WD that is merging with the first WD \citep[e.g.,][and references therein]{tornambe2005}.
The first possibility is the single-degenerate (SD) scenario and the second, the double-degenerate (DD) scenario.
In the SD scenario the accreted matter (which would be mostly hydrogen) burns to CO.
There are problems with both scenarios.
In the SD scenario, the accreted matter may be mostly ejected in some way because of unstable helium burning in the accreted matter 
and this prevents the WD central region 
from reaching explosive conditions (e.g., \citealt[and references therein]{piersanti2003};
\citealt[and references therein]{tornambe2005}).
In the DD scenario, it has been found that an off-center carbon ignition may lead to conversion of the CO WD to an O-Ne-Mg WD
\citep[e.g.,][]{saio2004} and perhaps, depending on conditions, to collapse to a neutron
star rather than a SN~Ia explosion \citep[e.g.,][]{nomoto1991}.
Somehow one of the scenarios or both of the scenarios avoid non-explosive fates and allow the CO WD to reach the central
conditions that lead to a SN~Ia explosion. 

    If a WD gets sufficiently close to the physical Chandrasekhar mass (and does not have
high rotation:  see below), it should collapse to a neutron star.
Electron capture on nuclei brings about collapse before the physical Chandrasekhar mass
is reached:  see Appendix~\ref{ap-chan-mass}.
SNe~Ia happen (at least in some cases) because of carbon ignition in the central regions of WDs (as mentioned above)
before the point of collapse is reached.

     In a SN~Ia explosion, nuclear burning turns the inner part of the ejecta into nearly pure IPEs with
the dominant isotope being radioactive $^{56}$Ni:  this inner part is the IPE core. 
Outside of the IPE core, IPEs are trace elements.
Immediately outside of the IPE core, the composition in most explosion models is dominated by
explosion-synthesized silicon and sulfur.
Above that is a layer dominated by explosion-synthesized and pre-existing oxygen with probably relatively
large abundances of explosion-synthesized silicon and magnesium.
There may be an outer layer or region of pre-existing CO that survives the burning in some SNe~Ia.
Carbon lines have been tentatively identified in a few SNe~Ia \citep[e.g.,][]{jeffery1992,fisher1999,branch2003}.

       The existence of a somewhat layered composition for SNe~Ia, including in particular an 
IPE core, seems almost certain.
Such compositions are the result of partially successful near-Chandrasekhar-mass, 
one-dimensional, hydrodynamic explosion models
\citep[e.g.,][]{nomoto1984,thielemann1986,woosley1991,woosley1994,khokhlov1993,hoeflich1996,hoeflich1998}
and are indispensable in modeling SNe~Ia with synthetic spectra.
A reference SN~Ia explosion model of longstanding exhibiting a layered composition is
the one-dimensional deflagration model~W7 \citep{nomoto1984,thielemann1986}.
In many respects model~W7 seems to approximate normal SN~Ia behavior to
some degree \citep[e.g.,][]{baron2006}\footnote{Plots of model~W7's density profile and composition, and data files that fully
specify model~W7 are available at
http://www.nhn.ou.edu/{\tt\~{}}jeffery/astro/sne/spectra/model/w7/w7.html\ .}.
Model~W7 does, in fact, have an outer layer of unburnt CO (with a trace of Ne) of $0.098\,M_{\odot}$
\citep{thielemann1986}.   
The unburnt mass fraction is $0.071$.
One-dimensional Delayed-detonation models, which have also been partially successful SN~Ia models, typically
have much less unburnt mass. 
For example, the near-Chandrasekhar-mass delayed-detonation models of 
\citet{khokhlov1993} have an unburnt mass fraction of order $0.01$ and those
of \citet{hoeflich1998} have an unburnt mass fraction of order $0.001$.
Based on this evidence, it is probable that normal SNe~Ia have unburnt mass fraction $\lesssim 0.07$.  

      A vital part of the standard SN~Ia model is the synthesis of the radioactive $^{56}$Ni in the explosion.
The $^{56}$Ni is the overwhelmingly dominant source of radiation energy for the observable phase of SNe~Ia.
The decay chain for  $^{56}$Ni is $^{56}$$\rm Ni\to^{56}$$\rm Co\to^{56}$$\rm Fe$
with \hbox{half-lives} of $6.077\pm0.012$~days and~$77.27\pm0.03$~days
for the first and second decays, respectively \citep[e.g.,][]{firestone}.

     H2006's analysis of the high luminosity of SN~2003fg led to the conclusion that SN~2003fg
had a very high mass of $^{56}$Ni and a (total) mass of $\sim 2\,M_{\odot}$, and thus that
SN~2003fg was super-Chandrasekhar (i.e., SNe~Ia with mass greater than the Chandrasekhar mass).
H2006 posit that the progenitor WD had high angular momentum, and thus a high centrifugal force.
The high centrifugal allowed the progenitor WD to reach 
a super-Chandrasekhar mass before explosive conditions were reached in the central regions. 
Both SD and DD scenarios may allow super-Chandrasekhar SNe~Ia. 
If the WD is spun up by accretion to very fast differential rotation (with mean angular velocity 
of order a few radians per second on average), then
the WD may exceed the physical Chandrasekhar mass by up to some tenths of a solar mass before reaching
explosive conditions in the central region \citep[hereafter YL]{yoon2005}.
The simple post-explosion super-Chandrasekhar SN~Ia models discussed in this paper (\S\S~\ref{section-mass} and~\ref{section-SSC}) are
independent of whether the pre-explosion evolution followed the SD scenario or DD scenario, and so we will not
discuss their distinctions further.
Before SN~2003fg, there was no strong evidence for super-Chandrasekhar SNe~Ia.

     In \S~\ref{section-mass}, we discuss the SN~2003fg $^{56}$Ni mass and (total) mass and H2006's analysis.
Section~\ref{section-IPE-core-boundary} discusses the SN~Ia IPE-core boundary and IPE-core velocity which
are relevant to our analysis of SN~2003fg.
Section~\ref{section-SSC} introduces our SN~Ia model (the SSC~model) for the analysis of the SN~2003fg $^{56}$Ni mass 
and mass.
In \S~\ref{section-comparison-sample}, we compare the predictions of the SSC~model to observations for a sample of low-$z$ SNe~Ia
from \citet{benetti2005} and in \S~\ref{section-comparison-sn2003fg}, to the observations for SN~2003fg.
Conclusions and discussion are given in \S~\ref{section-conclusions}.
Appendix~\ref{ap-chan-mass} discusses the meaning and values of the Chandrasekhar mass. 
Appendix~\ref{ap-YL-formula} presents the rotating CO WD binding energy formula of YL 
which H2006 used and which we use in the SSC~model and some related results for this binding energy formula.
Appendix~\ref{ap-SSC-mass} shows how to solve the SSC model for SN~Ia mass when SN~Ia mass is a dependent parameter.
Appendix~\ref{ap-SSC-IPE-core-mass} shows how to solve the SSC model for IPE-core mass when IPE-core mass is a dependent parameter.

\section{THE SN~2003fg $^{56}$Ni MASS AND TOTAL MASS\label{section-mass}}

     If we did not know that SN~2003fg was intrinsically exceptionally luminous, it would seem like
an only somewhat peculiar SNe~Ia and would attract only modest special attention.
But its luminosity and its violation of the lightcurve width-luminosity relationship makes it an exceptionally
interesting SN~Ia.   
In order to produce SN~2003fg's unusual luminosity,
H2006 have calculated that it had a mass of explosion-synthesized radioactive $^{56}$Ni of 
$M_{\rm ^{56}Ni}=1.29\pm0.07\,M_{\odot}$.

     The calculation for 
$M_{\rm ^{56}Ni}$ 
made use of the relationship
\begin{equation} 
           M_{\rm ^{56}Ni}={L_{\rm bol}\over \alpha \dot E_{\rm ^{56}Ni}(t_{\rm bol})} \,\, ,
\label{eq-alpha-factor}
\end{equation} 
where $L_{\rm bol}$ is the maximum of bolometric luminosity (which is emitted radiation integrated from
the ultraviolet to the infrared), 
$\dot E_{\rm ^{56}Ni}(t_{\rm bol})$ 
is the instantaneous rate
of release of radioactive decay energy per unit mass of $^{56}$Ni by
$^{56}$Ni and its daughter isotope $^{56}$Co at the time of maximum bolometric luminosity
$t_{\rm bol}$, and $\alpha$ is a correction factor to account
for time delay in the radioactive decay energy being emitted from the ejecta. 
H2006's value for 
$L_{\rm bol}/\dot E_{\rm ^{56}Ni}(t_{\rm bol})$ is $1.55\,M_{\odot}$.
From analytic solutions for the early SN~Ia lightcurves, \citet{arnett1979,arnett1982} 
found that $\alpha$ should be 1 exactly.   
In realistic SNe~Ia models and in real SNe~Ia, one would expect a multiplicity of interacting
effects will cause $\alpha$ to be only of order 1 and that there will be a range of values with
the value for any case depending on that case in perhaps a complex way. 
An early survey of SN~Ia lightcurve calculations suggested $\alpha=1.2\pm0.2$ \citep{branch1992}, where
the uncertainty represents the range of values.
An extensive set of SN~Ia lightcurve calculations \citep{hoeflich1996} gives $\alpha$ values that
extend over the range $0.71$--$1.46$ for models
that are at least partially successful in fitting lightcurves of one or more SNe~Ia:  the SNe~Ia included
normal SNe~Ia and peculiar SNe~Ia of various sorts.
The full $\alpha$ range (for models including those that are not acceptable for known SNe~Ia) is $0.62$--$1.46$
with mean $1.01$ and standard deviation $0.20$.
There are no unambiguous correlations of the $\alpha$ values with 
$M_{\rm ^{56}Ni}$,
rise time to bolometric maximum light, or model category \citep[Table~2, Fig.~7]{hoeflich1996}.
However, physically there should be a tendency for $\alpha$ to be relatively small if significant
$^{56}$Ni matter is near the surface which would happen necessarily if the $^{56}$Ni mass was a very
large fraction of the ejecta mass.
This tendency is because some of the radioactive decay gamma-rays can escape the ejecta directly from
near the surface, and thus not
deposit their energy in the ejecta:  such escaping gamma-rays are not included in the conventional
bolometric luminosity.
There is a hint of the gamma-ray-escape effect in the results of \citet{hoeflich1996}, but it is not clear how to identify
it among all the other interacting effects. 
    Actually most of the calculated lightcurves of \citet{hoeflich1996} have rise times to bolometric maximum
light that are too short for most SNe~Ia for which the mean rise time (using $B$ band as a proxy
for bolometric luminosity) is $\sim 19$~days \citep{conley2006}.
Nevertheless, the calculations of \citet{hoeflich1996} are suggestive of the range of $\alpha$ behavior that
may hold for SNe~Ia and we keep that in mind for our analysis in \S\S~\ref{section-comparison-sample}
and~\ref{section-comparison-sn2003fg}.
We also will keep in mind though that SN~2003fg is a very peculiar and a unique SN~Ia, and so its 
$\alpha$ value is not well constrained by the $\alpha$ values of existing SN~Ia models and other known SNe~Ia. 

     Following \citet{branch1992},
H2006 adopted $\alpha=1.2$ (which we will consider as the fiducial value)
for all their calculations of 
$M_{\rm ^{56}Ni}$, 
but did not consider
any variations of $\alpha$ nor include the uncertainty of $\alpha$ in their uncertainty calculations.
In fact, their other uncertainties (in bolometric luminosity $L_{\rm bol}$ and rise time to bolometric 
maximum light $t_{\rm bol}$) are negligible in comparison to the uncertainty in $\alpha$.

    Typically, SNe~Ia are thought to have of order $0.6\,M_{\odot}$ of $^{56}$Ni.
For the sample of well-observed 26 low-$z$ SNe~Ia studied by \citet{benetti2005} (hereafter the low-$z$ sample),
H2006 calculate a 
$M_{\rm ^{56}Ni}$ 
range of $\sim 0.04$--$0.86\,M_{\odot}$ (using $\alpha=1.2$).
Thus, H2006's value of 
$M_{\rm ^{56}Ni}=1.29\pm0.07\,M_{\odot}$ 
for SN~2003fg is well outside of the normal 
$M_{\rm ^{56}Ni}$ 
range and is the highest ever calculated with reasonable confidence
provided we assume SN~2003fg has no extreme variation in $\alpha$ from the fiducial value of $1.2$. 
(We will consider the possible variations in $\alpha$ and their effects on (total) SN~Ia mass estimates
in \S\S~\ref{section-comparison-sample} and~\ref{section-comparison-sn2003fg}.)
Note that the uncertainties of 
$L_{\rm bol}/\dot E_{\rm ^{56}Ni}(t_{\rm bol})$ 
for the low-$z$ sample are in
some cases greater than for SN~2003fg because some of 
these SNe~Ia are not well out in the Hubble flow where distances can be
determined with higher accuracy.
(Uncertain reddening corrections may also be a problem, but reddening corrections were not discussed by H2006.) 
Thus, the 
$M_{\rm ^{56}Ni}$ 
values for the low-$z$ sample are more uncertain than for SN~2003fg.

      The possibility of SNe~Ia with very high $^{56}$Ni mass has been raised in the past.
Assuming a long distance for the SN~1991T host galaxy NGC~4527 of $16.4\pm 1.0\,$Mpc,  
\citet{fisher1999} suggest a $^{56}$Ni mass of $\gtrsim 1.4\,M_{\odot}$ for SN~1991T.
However, Cepheid distances to NGC~4527 of $13.0\pm1.7\,$Mpc \citep{gibson2001} and
$14.1\pm0.8\,$Mpc \citep{saha2001} imply a less luminous SN~1991T, and thus a lower $^{56}$Ni mass that is in or
closer to the normal range. 
For SN~1991T, H2006 find 
$M_{\rm ^{56}Ni}=0.86\,M_{\odot}$
using $\alpha=1.2$:  it is not
clear what distance they adopted for NGC~4527.  
Several SNe~Ia or possible SNe~Ia have very uncertain absolute $B$ maxima  \citep{richardson2002}
that are about as bright or even brighter than that of SN~2003fg.
The most outstanding example is possible SN~Ia SN~1988O for which the brightest observed absolute $B$ magnitude
is $\sim -21.3$ (using $H_{0}=70\,{\rm km\,s^{-1}\,Mpc^{-1}}$), but probably with great uncertainty
\citep{richardson2002}.
This $B$ value is $\sim 1.2$ magnitudes brighter than SN~2003fg's absolute $B$ maximum 
of $-20.09$ (H2006).  
The uncertainties are such that there is no confidence at present that any of the possibly
very bright SNe~Ia are as bright or have $^{56}$Ni mass comparable to SN~2003fg.  

     The high $^{56}$Ni mass of SN~2003fg has dramatic implications for SN~2003fg's (total) mass. 
As discussed in \S~\ref{section-introduction}, in the standard SN~Ia model, a SN~Ia has only
about a physical Chandrasekhar mass $1.38\,M_{\odot}$ since that is the standard-SN-Ia-model mass of its WD progenitor.
However, accreting, rotating WDs can exceed the Chandrasekhar mass before reaching the density and temperature
conditions for explosion.
The centrifugal force provides the WD with extra support against gravity.
YL believe that WDs of up to $\sim 4\,M_{\odot}$ are dynamically possible, but
that possible accretion histories may limit them to not much more than $2\,M_{\odot}$ \citep{langer2000}.  

     Given that the calculated SN~2003fg $^{56}$Ni mass is so close to the Chandrasekhar mass, it is difficult on simple
grounds to believe that SN~2003fg's mass is only nearly a Chandrasekhar mass.
If SN~2003fg were a near-Chandrasekhar-mass SN~Ia and its radiative transfer history not exotic, then its photosphere
would be well inside its IPE core near $B$ maximum for density profiles that work well for normal SNe~Ia.
An IPE core of $1.29\,M_{\odot}$ of $^{56}$Ni and nothing else (which is implausible)
in a near-Chandrasekhar-mass SN~2003fg with a model-W7 density profile would extend to $\sim 15000\,{\rm km\,s^{-1}}$
\citep[Fig.~4]{nomoto1984}.
If this were the case, the SN~2003fg 2-days-past-$B$-maximum spectrum would be dominated by IPE lines and, as
discussed in \S~\ref{section-introduction}, this is not observed.
To keep the  2-days-past-$B$-maximum spectrum looking normal, the actual SN~2003fg IPE-core velocities must probably have been
kept below $\sim 8000\,{\rm km\,s^{-1}}$ (the SN~2003fg photospheric velocity for 2~days past $B$~maximum:  \S~{section-intro}) 
by much more than $0.1\,M_{\odot}$ of overlying mass.

     A simple scaling argument gives a possible mass for SN~2003fg.
Assume that the ratios of $^{56}$Ni mass 
$M_{\rm ^{56}Ni}$
and other elements to mass $M$ 
are roughly constant for SNe~Ia of any mass. 
Then mass will roughly scale with $^{56}$Ni mass.
If the average SN~Ia has $0.6\,M_{\odot}$ of $^{56}$Ni and 
a near-Chandrasekhar mass, then SN~2003fg would have a mass of
\begin{equation}
               M\approx {1.29\over 0.6/1.4}=3.0 \,M_{\odot} \,\, . 
\end{equation}
If we now assume that $\sqrt{2E/M}$ is approximately a constant as mass is varied, 
the ejecta velocities for the various matter elements of SNe~Ia will be crudely invariant 
with mass since velocities for the various matter elements overall scale approximately as $\sqrt{2E/M}$.
Thus, a $3\,M_{\odot}$ SN~2003fg could have the velocities of line formation near $B$ maximum that
were not wildly different from those of normal SNe~Ia provided the photospheric velocity evolution
was not wildly different from that of normal SNe~Ia.
The assumptions we made to obtain the estimate of $3\,M_{\odot}$ for SN~2003fg are at most
only partially valid, and thus the estimate is only suggestive.
We verify in Appendix~\ref{ap-YL-formula} that $\sqrt{2E/M}$ can be crudely accepted as constant with mass $M$ provided
super-Chandrasekhar SNe~Ia are approximated by our SSC model (see \S~\ref{section-SSC}) up to masses of $3\,M_{\odot}$.

     Another estimate of the SN~2003fg mass follows from an argument given by H2006
that we modify a bit here.      
If an average SN~Ia has a near-Chandrasekhar mass and 
$M_{\rm ^{56}Ni}=0.6\,M_{\odot}$, 
then it has about $0.8\,M_{\odot}$ of non-$^{56}$Ni matter.
If SN~2003fg is to look as spectroscopically normal as it does, then perhaps it should have of
order the same amount of non-$^{56}$Ni matter.
It follows then that the SN~2003fg mass could be $\sim 2.1\,M_{\odot}$.
This mass estimate is also only suggestive.

    H2006 also estimate the SN~2003fg mass from a simple SN~Ia model that allows for super-Chandrasekhar
mass since it assumes a rotating WD progenitor and that has
4 independent parameters:  (total) mass $M$, mass of explosion-synthesized IPEs $M_{\rm IPE\mathhyphen core}$, 
surviving unburnt CO mass fraction $f_{\rm CO}$,
and 
progenitor WD central density $\rho_{\rm central,WD}$.
H2006 apparently in all cases set $\rho_{\rm central,WD}=4\times10^{9}\,{\rm g\,cm^{-3}}$ which can
be regarded as of order a typical progenitor WD central density.    
We use the symbol $M_{\rm IPE\mathhyphen core}$ for the explosion-synthesized IPE mass since
for the SSC~model (introduced in \S~\ref{section-SSC}), we will make the assumption
that all the explosion-synthesized IPEs are in the IPE core and no other matter is in the IPE core.
H2006 do not, in fact, make this assumption although it does not change their model much to do so. 
For the sake of consistent notation, we will use $M_{\rm IPE\mathhyphen core}$ for the description of H2006's model.
The mass fractions of the ejecta of 
explosion-synthesized IMEs and explosion-synthesized IPEs 
are, respectively, 
$f_{\rm IME}$, and $f_{\rm IPE}$.
(Note carbon and oxygen are IMEs, but the unburnt amount of CO is not included in $f_{\rm IME}$.) 
These are not independent parameters since
\begin{equation}
   f_{\rm IPE}={M_{\rm IPE\mathhyphen core}\over M}        \qquad\hbox{and}\qquad
   f_{\rm IME}=1-f_{\rm CO}-f_{\rm IPE} \,\, .
\label{eq-fipe-fime}
\end{equation}

     Actually, to obtain model parameter $M_{\rm IPE\mathhyphen core}$ from 
$M_{\rm ^{56}Ni}$, 
one needs a conversion factor
\begin{equation}
               g={ M_{\rm ^{56}Ni}\over M_{\rm IPE\mathhyphen core} } \,\, .
\label{eq-g-factor}
\end{equation}
H2006 adopt $g=0.7$.
The rationale for this value is as follows.
The deflagration model~W7 gives $g=0.70$   
\citep{nomoto1984,thielemann1986} (although to find this number accurately one must examine the model~W7 data files)
and the pure detonation model DET1 (which is not a reasonable model for any SN~Ia) gives $g=0.66$  
\citep{khokhlov1993}.  
Models~W7 and DET1 are rather remote from each other in model space, and so the fact that they yield similar $g$ values
near $0.7$ suggests that $g=0.7$ is a good representative fiducial value.
Obviously $g$ must be in the range 0 to 1.

     From equations~(\ref{eq-alpha-factor}) and~(\ref{eq-g-factor}), the full formula for $M_{\rm IPE\mathhyphen core}$ is
\begin{equation}
               M_{\rm IPE\mathhyphen core}={L_{\rm bol}\over g\alpha \dot E_{\rm ^{56}Ni}(t_{\rm bol}) } \,\, .
\label{eq-IPE-core-mass}
\end{equation}
As mentioned above, we believe that the uncertainties in the SN~2003fg $L_{\rm bol}$ and $t_{\rm bol}$ are negligible
in comparison to the uncertainty of $\alpha$ for SN~2003fg.
The uncertainty in $g$ for SN~2003fg may be smaller than that of $\alpha$, but it is still not very well
constrained.
In our analysis in
\S\S~\ref{section-comparison-sample} and~\ref{section-comparison-sn2003fg},
we will consider how the results change when varying
$\alpha$ and $g$ and neglect uncertainties in $L_{\rm bol}$ and $t_{\rm bol}$.

    The kinetic energy of H2006's SN~Ia model is the energy released by nuclear burning minus the binding energy of
the WD progenitor and is assumed to be given by 
\begin{equation}
         E=E_{\rm IPE}M[f_{\rm IPE}+R(1-f_{\rm CO}-f_{\rm IPE})] - E_{\rm bind}(\rho_{\rm central,WD},M)  \,\, ,
\label{eq-energy-a}
\end{equation}
where burning coefficient $E_{\rm IPE}=1.61\,{\rm foe}/M_{\odot}$ (1~foe=$10^{51}\,{\rm ergs}$) is the energy per unit
mass of burning CO (with equal amounts of carbon and oxygen) to IPEs, 
$M$ is the WD/supernova mass in solar mass units, 
$R=0.768$ is the fraction of $E_{\rm IPE}$ 
released by burning CO (with equal amounts of carbon and oxygen) to IMEs, 
$(1-f_{\rm CO}-f_{\rm IPE})=f_{\rm IME}$ from equation~(\ref{eq-fipe-fime}),
$E_{\rm bind}(\rho_{\rm central,WD},M)$ is the binding energy of the progenitor WD,
and 
$\rho_{\rm central,WD}$ is the central density of the progenitor WD.
The value of $E_{\rm IPE}$ is weighted average of the burning coefficients for burning 
CO (with equal amounts of carbon and oxygen) to $^{56}$Ni ($1.56\,{\rm foe}/M_{\odot}$) and an average
of stable IPEs ($1.74\,{\rm foe}/M_{\odot}$).
The weights for the averaging are $0.7$ for $^{56}$Ni and $0.3$ for the stable IPEs:   these
weights agree with the our adoption of fiducial $g=0.7$ (see above). 
The value of $R$ is obtained by dividing an averaged energy per unit
mass of burning CO (with equal amounts of carbon and oxygen) to IMEs
($1.24\,{\rm foe}/M_{\odot}$) by $1.61\,{\rm foe}/M_{\odot}$.
The burning coefficient values $1.56\,{\rm foe}/M_{\odot}$, $1.74\,{\rm foe}/M_{\odot}$, and
$1.24\,{\rm foe}/M_{\odot}$ were taken from \citet{woosley2006}.
(Actually, used $E_{\rm IPE}=1.55\,{\rm foe}/M_{\odot}$ and $R=0.76$ which are older
value given by \citet{branch1992}.
The difference between the older and newer values is negligible for H2006's analysis.)

    The binding energy formula in equation~(\ref{eq-energy-a}) comes from YL 
and applies to WDs (including super-Chandrasekhar-mass WDs) partially sustained against collapse by rotation.
The formula is based on YL's theory of differential rotation of mass-accreting WDs and
is for zero temperature WDs.
The effects of finite temperature are small for the purposes of H2006 and the SSC~model
(which also uses eq.~(\ref{eq-energy-a}): see \S~\ref{section-SSC} and Appendix~\ref{ap-chan-mass}).
The mean angular velocities of super-Chandrasekhar-mass WD models YL consider are typically of order 
a few radians per second (see YL's Tables~1 and ~2).
The formula is verified for the $\rho_{\rm central,WD}$ range $\sim 10^{8}$---$10^{10}\,{\rm g\,cm^{-3}}$
and
the mass range $\sim 1.16$--$2.05\,M_{\odot}$. 
YL's formula and rotating super-Chandrasekhar-mass WD models were introduced with the SD scenario as the application
of primary interest, but they should also apply to the DD scenario if the WD in that scenario avoids off-center
carbon ignition \citep{yoon2006}. 
We present YL's binding energy formula and ancillary formulae
for reference in Appendix~\ref{ap-YL-formula}.               

    The $\rho_{\rm central,WD}$ range for YL's binding formula includes the range that can lead to WD explosion. 
No explosion probably happens for $\rho_{\rm central,WD}\lesssim 2.5\times10^{9}\,{\rm g\,cm^{-3}}$ \citep[e.g.,][]{woosley2006}
and
above $\rho_{\rm central,WD}\approx10^{10}\,{\rm g\,cm^{-3}}$
the WD will probably experience electron-capture induced collapse to a neutron star for plausible accretion histories
rather than explode
(\citealt{nomoto1991};  YL).  
\citet{woosley2006} adopt $\rho_{\rm central,WD}=2.9\times10^{9}\,{\rm g\,cm^{-3}}$ as their
fiducial explosion density based on their explosion modeling experience:  we follow them in adopting
this density as the fiducial explosion density for most purposes in this paper. 
The  $E_{\rm bind}(\rho_{\rm central,WD},M)$ is approximately linear in mass for a fixed
$\rho_{\rm central,WD}$ (see YL's Fig.~9). 
For each $\rho_{\rm central,WD}$, there are minimum $M$ and $E_{\rm bind}$
which are the mass and binding energy of a non-rotating WD.
The minimum $M$ and $E_{\rm bind}$ grow slightly with $\rho_{\rm central,WD}$ for the 
range $2.5\times10^{9}$---$10^{10}\,{\rm g\,cm^{-3}}$:
mass from $1.384\,M_{\odot}$ to $1.414\,M_{\odot}$;
binding energy from $0.501\,$foe to $0.570\,$foe (see 
eq.~(\ref{eq-mass-nr}) and~(\ref{eq-energy-bind-nr}) in Appendix~\ref{ap-YL-formula}). 
Note that the non-rotating mass range is above the physical Chandrasekhar mass $1.38\,M_{\odot}$.
This is because YL use an approximate equation of state for their modeling that gives a Chandrasekhar
mass of $1.436\,M_{\odot}$ (see Appendices~\ref{ap-chan-mass} and~\ref{ap-YL-formula}).
The inconsistency with the physical Chandrasekhar mass is small in comparison to the mass variations
H2006 and we consider. 

   YL's binding energy formula is, in fact, only a close fit to numerical simulations.
Those simulations rely on many approximations and YL's theory of differential rotation in
mass-accreting WDs.
So YL's binding energy formula is not the last word in binding energy formulae.
Quantitatively, the binding energy formula is important. 
If one sets $f_{\rm CO}=0.0$ and $f_{\rm IPE}=0.5$, which are somewhat realistic for SNe~Ia, then
the nuclear burning energy part of equation~(\ref{eq-energy-a}) becomes 
\begin{equation}
   E_{\rm nuc}=1.43\times (M-1.4)+2.00 \; \hbox{foe}  \,\, ,
\label{eq-nuclear}
\end{equation}
where $M$ is solar masses and we have used $M=1.4\,M_{\odot}$ as a fiducial value for this argument since
that is approximately the minimum mass for explosion.
The binding energy formula can be approximated as a line to high accuracy (Appendix~\ref{ap-YL-formula}) with
coefficients depending on density.
For our fiducial $\rho_{\rm central,WD}=2.9\times10^{9}\,{\rm g\,cm^{-3}}$,
the binding energy formula is
approximated for the mass range $\sim 1.4$--$3\,M_{\odot}$ (which turns out to be the mass range of interest in this paper)
by the least-squares-fit line
\begin{equation}
   E_{\rm bind}\approx 1.220\times (M-1.4)+0.506 \; \hbox{foe}  \,\, ,  
\label{eq-bind-approx}
\end{equation}
where $M$ is solar masses and we have again used $M=1.4\,M_{\odot}$ as a fiducial value.
Comparing equations~(\ref{eq-nuclear}) and~(\ref{eq-bind-approx}), it is clear that the binding energy will always 
be important to the behavior of the ejecta for the mass range of interest in this
paper (which, again, is $\sim 1.4$--$3\,M_{\odot}$)
since the nuclear burning energy is never overwhelmingly dominant for this range.
Thus, the quantitative results of H2006's SN~Ia model and the SSC model do depend on the YL's binding energy formula.
But it is likely that improvements to YL's formula will not change these results qualitatively, unless 
rotating WDs with masses significantly greater than the Chandrasekhar mass are ruled out by improvements
in the input physics of YL's theory of differential rotation in mass-accreting WDs. 

     Using the kinetic energy from equation~(\ref{eq-energy-a}) with $\rho_{\rm central,WD}=4\times10^{9}\,{\rm g\,cm^{-3}}$
in all cases, H2006 calculate characteristic kinetic-energy velocities using the formula
\begin{equation}
   v_{\rm kin}=\sqrt{2E \over M} = 10027\,{\rm km\,s^{-1}}\times   
                    \sqrt{ \left({E\over 1\,{\rm foe} }\right)
                           \left({\,M_{\odot} \over M}\right)}         \,\, ,
\label{eq-kinetic-energy-velocity}
\end{equation}
where the second expression was created using $M_{\odot}=1.9891\times10^{33}\,{\rm g}$ \citep[e.g.,][p.~12]{cox2000}.
H2006's Figure~4b--c shows curves of $v_{\rm kin}$ versus $^{56}$Ni mass 
($M_{\rm ^{56}Ni}$) 
where the curves are for WD masses in the range from $1.4\,M_{\odot}$ (i.e., approximately 
the Chandrasekhar mass:  see Appendix~\ref{ap-chan-mass}) to $2.1\,M_{\odot}$ and a range of $f_{\rm CO}$ values from $0$ to $0.3$.

    For an observation-derived quantity to compare to the $v_{\rm kin}$ curves, H2006 make use of the 26
SNe~Ia of the low-$z$ sample.
On their Figure~4b--c, H2006 plot Si~II~$\lambda6355$ line velocities for the low-$z$ sample
for 40 days past $B$ maximum calculated from a least-squares fit
to the line velocity evolutions which in some cases extend to 40 days or a bit more past $B$ maximum  \citep{benetti2005}.
If the composition change between the IPE core and silicon-sulfur layer above
the IPE core is reasonably sharp (as explosion models suggest:  e.g., model~W7), then the Si~II~$\lambda6355$ line velocity
of a SNe~Ia should stop decreasing (we can say it plateaus) at about the velocity of the IPE-core boundary (which we call the
IPE-core velocity for brevity) since the line formation for silicon and other IME P~Cygni lines cannot 
recede into the IPE core where there is almost no IME matter to cause IME line formation.
This plateauing of the Si~II~$\lambda6355$ line velocity may well have happened in some of the low-$z$ sample 
as is suggest by the data from about 40 days past $B$ maximum \citep[Fig.~1]{benetti2005}.
Unfortunately, line blending with iron lines that develop after $B$ maximum may change the
Si~II~$\lambda6355$ line velocity away from being just the IPE-core velocity. 
Nevertheless, the long-post-$B$-maximum Si~II~$\lambda6355$ line velocities (when the Si~II~$\lambda6355$ line
can be identified) may be a good approximation to the IPE-core velocity.
We discuss determination of the IPE-core velocities further in \S~\ref{section-IPE-core-boundary}. 

      H2006 use the slope of the best fit line of that member of the low-$z$ sample that
is closest to SN~2003fg on a plot of near-$B$-maximum Si~II~$\lambda6355$ line velocity versus
$M_{\rm ^{56}Ni}$
(H2006's Fig.~4a) to extrapolate from
their SN~2003fg 2-day-past-$B$-maximum Si~II~$\lambda6355$ line velocity
to a 40-days-past-$B$-maximum Si~II~$\lambda6355$ line velocity for SN~2003fg.
They plot this on their Figure~4b--c along with the low-$z$ sample 40-days-past-$B$-maximum Si~II~$\lambda6355$ line velocities.
Taking their Figure~4b--c at face value, except for SN~2003fg, all the SNe~Ia from 
the low-$z$ sample fall in their figure in the range where a near-Chandrasekhar mass
is favored or cannot be excluded if $f_{\rm CO}$ is high.
(The higher $f_{\rm CO}$, the lower $f_{\rm IME}$ low) and the less energy from nuclear burning, 
and so the lower characteristic kinetic-energy velocity
without having to invoke extra mass beyond the Chandrasekhar mass to slow the ejecta down.)
If $f_{\rm CO}$ is low (or $f_{\rm IME}$ high), some of these SNe~Ia may be super-Chandrasekhar with masses up to $\sim 2\,M_{\odot}$
again taking H2006's Figure~4b--c at face value.

     SN~2003fg is well away from the low-$z$ sample and is clearly in the super-Chandrasekhar mass region with
mass $\gtrsim 2\,M_{\odot}$ for $f_{\rm CO}\lesssim 0.2$.  
As discussed in \S~\ref{section-introduction}, it is probable that normal SNe~Ia have $f_{\rm CO}\lesssim 0.07$ and,
in the context of energy production from the nuclear burning of CO, $0.07$ is much less than $0.2$. 
The spectrum of SN~2003fg does show a possible multiplet C~II~$\lambda4267$ P~Cygni line 
\citep[e.g.,][p.~39]{wiese1966} 
that has hitherto been only very tentatively identified in SN~Ia spectra \citep{branch2003}
and would be expected to be
weaker in local-thermodynamic equilibrium (LTE) than C~II lines that have been tentatively identified in SN~Ia spectra. 
Those other C~II lines may be hidden in the noise of the SN~2003fg spectra.
If the C~II line identification is correct, it suggests SN~2003fg may have a larger than normal unburnt
CO abundance in its outer layers.
But given the overall normality of the SN~2003fg spectrum, $f_{\rm CO}\lesssim 0.2$ is probable.
Thus, from H2006's Figure~4b--c,
one would find the SN~2003fg mass to be $\gtrsim 2\,M_{\odot}$.
Given the conclusion of \citet{langer2000} that WD masses much greater than $\sim 2\,M_{\odot}$ are not attainable
in possible accretion histories, H2006's Figure~4b--c suggests a SN~2003fg mass of $\sim 2\,M_{\odot}$.

     Unfortunately, as H2006 emphasize 
the comparison of 40-days-past-$B$-maximum Si~II~$\lambda6355$ line velocities 
and characteristic kinetic-energy velocity is ambiguous.
The characteristic kinetic-energy velocity has no definite theoretical relation to 
the line velocities or the IPE-core velocities that the line velocities probably approximate. 
Thus, the comparison on their Figure~4b--c must be regarded as only suggestive. 
But the fact that SN~2003fg is so far away from the other SNe~Ia on the plot and well into the 
super-Chandrasekhar mass region is striking.
     
       Taken altogether the arguments above give reasonable evidence that SN~2003fg had a super-Chandrasekhar mass
which is perhaps $\sim 2\,M_{\odot}$. 
One would like, however, a more reliable quantitative value for the SN~2003fg mass or at least more
robust lower bounds on this mass.
In \S~\ref{section-SSC}, we introduce a simple model (the SSC~model) which is an improvement on the model of H2006 and
which gives lower bounds on the SN~2003fg mass of varying confidence.

\section{THE IPE-CORE BOUNDARY AND THE IPE-CORE VELOCITY\label{section-IPE-core-boundary}}

     Before going on to the SSC~model, some further discussion of IPE-core boundary and IPE-core
velocity (see \S~\ref{section-mass}) needs to be made    
to support the validity of approximating the IPE-core boundary as perfectly
sharp in the SSC~model (see \S~\ref{section-SSC}) and to relate the IPE-core velocity to 
more directly observable quantities. 

      The existence of a relatively sharp boundary for the IPE core for normal SNe~Ia at least
is basically supported by spectrum modeling using explosion models. 
Typically in useful SN~Ia explosion models, the IPEs fall from dominating the composition
(nearly $100\,$\% of the composition by mass)
to being a trace (a few percent by mass) over a few thousands of kilometers per second in velocity coordinate.
For example, the dominance-to-trace transition in the partially successful
models W7 \citep{nomoto1984,thielemann1986} and DD4
(a well-known delayed-detonation model)
\citep{woosley1991} happens in the range $\sim 8500$--$11000\,{\rm km\,s^{-1}}$.
The overall SN~Ia ejecta velocity scale is $\sim 25000\,{\rm km\,s^{-1}}$ or more, and
so assuming a perfectly sharp boundary is reasonable in a simple model.
We must add, however, that some mixing or explosion-synthesis of IPEs beyond the IPE-cores of explosion
models is sometimes helpful in SN~Ia spectrum modeling \citep{baron2006}.
We will not consider this complication further in this paper it should be kept in mind for future work. 

     Since the IPE-core boundary is not perfectly sharp, one needs a definition for the characteristic
IPE-core velocity.
Probably the most sensible definition of the characteristic IPE-core velocity for explosion models is the velocity of the spherical
shell that would just contain all the explosion-synthesized IPEs and nothing else given the model comoving density profile. 
For model~W7, this velocity is $9800\,{\rm km\,s^{-1}}$ which just $\sim 200\,{\rm km\,s^{-1}}$ less than 
the velocity where the mass fractions of $^{56}$Ni and silicon are about equal.   
For a quick survey, we therefore adopt the velocity of equality of $^{56}$Ni and silicon mass
fractions as an adequate proxy for the IPE-core velocity.
For some models that have been found useful in analyzing SNe~Ia, the equality velocities
fall in the range $\sim 7000$--$13000\,{\rm km\,s^{-1}}$ 
\citep[e.g.,][]{nomoto1984,thielemann1986,woosley1991,khokhlov1993,hoeflich1998}.   
For the low-$z$ sample, most of the calculated Si~II~$\lambda6355$ line velocities for 40 days 
past $B$ maximum are in the range
$\sim 8000$--$11000\,{\rm km\,s^{-1}}$ either by measurement or extrapolation:
the five faint SNe~Ia have lower Si~II~$\lambda6355$ line velocities
in the range $\sim 5000$--$8000\,{\rm km\,s^{-1}}$ either by measurement or extrapolation
\citep{benetti2005}.
(Faint SN~Ia is a classification of citet{benetti} for those SNe~Ia that are distinctly fainter than
normal SNe~Ia:  the prototypes of this classification are SN~1986G and SN~1991bg.)
The 40-day-past-$B$-maximum values for the non-faint SNe~Ia are roughly consistent 
with equality velocities we cite and support the idea that
the long-post-$B$-maximum Si~II~$\lambda6355$ line velocities at least for non-faint SNe~Ia 
approximate the IPE-core velocities.
Recall that some of these 40-day-past-$B$-maximum Si~II~$\lambda6355$ line velocities seem to
be plateauing \citep[Fig.~1]{benetti2005}. 

      Besides the Si~II~$\lambda6355$ line, other IME P~Cygni lines are useful in locating or at least
constraining the IPE-core velocity. 
For example, the resonance multiplet Na~I~$\lambda5892$ and calcium multiplets Ca~II $\lambda3945$ (resonance)
and Ca~II $\lambda8579$ (non-resonance, but arising from a metastable level) 
\citep[e.g.,][p.~2, 252, 251]{wiese1969}
give rise to P~Cygni profiles that can persist to 
100~days or more past $B$~maximum and, in the core normal SN~Ia SN~1994D \citep[e.g.,][]{branch2005} for example, have 
line velocities $\sim 11000\,{\rm km\,s^{-1}}$, $\sim 8000\,{\rm km\,s^{-1}}$, and
$\sim 10000\,{\rm km\,s^{-1}}$, respectively, 
at 115 days past $B$~maximum.    
In the ultraviolet, the P~Cygni line of resonance multiplet Mg~II $\lambda2798$ \citep[e.g.,][p.~30]{wiese1969}
also seems long persistent
\citep{kirshner1993,ruiz-lapuente1995} with a line velocity that seems to stay $\gtrsim 14000\,{\rm km\,s^{-1}}$
to 291 days past $B$ maximum although line-blending with IPE lines
makes it hard to be sure that this velocity is representative of the magnesium layer. 
The Mg~II $\lambda2797.9$ region of the ultraviolet is rarely observed for low-$z$ supernovae (because the
observation requires space-based telescopes which are not always available), but for high-$z$ ones
the Mg~II $\lambda2797.9$ region can be redshifted into more readily observed wavelength ranges. 
H2006's spectrum of SN~2003fg almost extends far enough to the blue to see the region of the 
Mg~II $\lambda2797.9$ line.
The late, persistent IME P~Cygni lines with relatively high line velocities suggest that the line formation layers
for these lines has receded to the vicinity of or some point above the IPE core and then plateaued since
there is almost no IME matter in the IPE core for lower velocity IME line formation.
Thus, the sodium, magnesium, and calcium line velocities may constrain the IPE-core velocity. 
On the theoretical side,
explosion models suggest that the abundances of sodium, magnesium, and calcium all decline going inward in
the ejecta.
For model~W7, sodium, magnesium, and calcium
fall below their solar mass fractions $3.2\times10^{-5}$, $6.0\times10^{-4}$, and $5.7\times10^{-5}$,  
respectively, \citep{asplund2005}
at velocities $\sim 14500\,{\rm km\,s^{-1}}$, $\sim 13000\,{\rm km\,s^{-1}}$, and
$\sim 9000\,{\rm km\,s^{-1}}$, respectively.
These velocities are only representative, of course, but they also suggest that the lowest observed
line velocities for Na~I~$\lambda5892$, Mg~II $\lambda2798$, Ca~II $\lambda3945$, Ca~II $\lambda8579$ for a SN~Ia
could constrain the IPE-core velocity.
Recall from above, that for model~W7 the characteristic IPE-core velocity is $\sim 9800\,{\rm km\,s^{-1}}$.

     Besides P~Cygni lines, one can also consider for the determination of the IPE-core boundary
the forbidden Fe~II and Fe~III emission lines
that dominate the optical SN~Ia spectrum in the nebular phase when the ejecta become optically thin.
Modeling these emission lines can be used to constrain the IPE-core velocity.
Current results suggest IPE-core velocities in the range $\sim 7500$--$10500\,{\rm km\,s^{-1}}$
for normal and somewhat bright SNe~Ia and IPE-core velocities of $\sim 2000\,{\rm km\,s^{-1}}$ for
very faint SNe~Ia like SN~1991bg \citep{mazzali1998}. 
Note that the calculations for these IPE-core velocities apparently assumed a uniform density for the IPE core
(as by \citet{mazzali1997}) which a significant approximation. 

     The above discussion shows both from explosion models and observations that something 
like a sharp IPE-core boundary does exist, a characteristic IPE-core velocity can
be defined, and that this  characteristic IPE-core velocity is in practice determinable
to within some uncertainty.

\section{THE SSC MODEL FOR SUPER-CHANDRASEKHAR-MASS SNe~Ia\label{section-SSC}}

       In this section, we introduce a  model  
for studying super-Chandrasekhar-mass SNe~Ia that is an improvement on the model of H2006.
We call this model the SSC model for Simple Super-Chandrasekhar model for SNe~Ia.
The progenitor for the model is, as in the standard SN~Ia model, a mass-accreting CO WD.
The SSC model has 3 components.
The first component is equation~(\ref{eq-energy-a}) (\S~\ref{section-mass})
(slightly rewritten:  see below) to determine the total energy of the SN~Ia ejecta.
Because we use this equation, we assume we can approximate ejecta as consisting of only three types of
material for the kinetic energy calculation:  
explosion-synthesized IPEs, explosion-synthesized IMEs, and unburnt CO. 
By using equation~(\ref{eq-energy-a}), we are, of course, also assuming that the progenitor WD 
is rotating for masses significantly greater than the physical Chandrasekhar mass $1.38\,M_{\odot}$
(see Appendix~\ref{ap-chan-mass}) and we are relying on YL's binding energy formula for rotating WDs.
As discussed in \S~\ref{section-mass} our results from the SSC~model quantitatively depend on 
YL's binding energy formula.
Qualitatively, the results from the SSC~model are probably robust.  

       The second component of the SSC~model is the exponential model of the ejecta 
\citep{jeffery1999} which is explained as follows.
Recall from \S~\ref{section-introduction} that after very early times supernovae are in homologous expansion
where velocity becomes a good comoving frame coordinate and all mass element densities scale as $t^{-3}$. 
Now as has been known for some time, many partially successful SN~Ia models (e.g., model~W7)
have comoving density profiles that are approximately inverse exponentials of velocity, \citep{jeffery1992}.
(Even earlier, it was recognized that the comoving density profiles of core-collapse supernova models were
sometimes approximately piecewise inverse exponentials with velocity \citep{pizzochero1990}.)
Making use of homologous expansion and the (inverse) exponential density profile approximation,
one obtains a useful parameterization of approximate SN~Ia density profiles: 
\begin{equation}
    \rho(v,t)=\rho_{\rm central,0}\left({t_{0}\over t}\right)^{3}\exp\left(-{v\over v_{e}}\right) \,\, ,
\end{equation}
where $t_{0}$ is some fiducial time, $\rho_{\rm central,0}$ is the central density of the ejecta
at the fiducial time,
and
$v_{e}$ is the $e$-folding velocity.
For examples of SN~Ia model profiles that
approximate exponentials see, e.g., \citet[Fig.~1]{dwarkadas1998}, \citet[Fig.~11]{baron2006}, and \citet[Fig.~4]{woosley2006}.
It is straightforward to show that
\begin{equation}
         v_{e}=\sqrt{{1\over6}{E\over M}}
                   =2894.7\,{\rm km\,s^{-1}}\times      
                    \sqrt{ \left({E\over 1\,{\rm foe} }\right)
                           \left({\,M_{\odot} \over M}\right)}         \,\, ,
\label{eq-efolding}
\end{equation}
where, as in \S~\ref{section-mass}, $E$ is the ejecta kinetic energy and $M$ is the ejecta mass
\citep{jeffery1999}. 
Since SN~Ia kinetic energy should be of order a foe and mass of order $M_{\odot}$, the coefficient of
the second expression of equation~(\ref{eq-efolding}) can be take as a good fiducial value for
the SN~Ia $e$-folding velocities.
To support this we can consider model~W7.
Using the $E$ and $M$ values for model~W7 in equation~(\ref{eq-efolding}) gives $v_{e}=2670\,{\rm km\,s^{-1}}$
which is not far from $2894.7\,{\rm km\,s^{-1}}$ and
which, in fact, leads to a good fit to the model~W7 density profile:  see the URL given in footnote~2 for plots
showing this fit. 
We note that the characteristic kinetic-energy velocity defined by 
equation~(\ref{eq-kinetic-energy-velocity}) (\S~\ref{section-mass}) is related to
$v_{e}$ by 
\begin{equation}
 v_{\rm kin}=\sqrt{12} v_{e}  \,\, .
\end{equation}
It is also straightforward to show that 
the fraction of ejecta mass, $f$, interior to reduced velocity 
\begin{equation}
x={v\over v_{e}}
\label{eq-reduced-velocity}
\end{equation}
is given by 
\begin{equation}
f=
\cases{\displaystyle
  \left[1-\exp(-x)\left(1+x+{1\over2}x^{2}\right)\right]   &in general;  \cr
\noalign{\smallskip}
\displaystyle
   \left\{1-\exp(-x)\left[\exp(x)-\sum_{k=3}^{\infty}{x^{k}\over k!}\right]\right\} &in general; \cr
\noalign{\smallskip}
\displaystyle
    {x^{3}\over6}-{x^{4}\over8}+{x^{5}\over20}     &to 5th order in small $x$; \cr  
\noalign{\smallskip}
    0.080301397071394\ldots   &for $x=1$; \cr    
\noalign{\smallskip}
\displaystyle
    0.67246197033808\ldots \approx {2\over3}    &for $x=\sqrt{12}=3.4641016151377\ldots$; \cr
\noalign{\smallskip}
    0.99723060428448\ldots                     &for $x=10$ \cr}
\label{eq-interior-mass-fraction}
\end{equation}
\citep{jeffery1999}.

       Note that the exponential density profile extends to infinite velocity, in principal.
Obviously, the approximation of an exponential density profile must fail as velocity becomes relativistic.
For modeling SNe~Ia, this is not a problem since there is seldom need to invoke
matter beyond $\sim 30000\,{\rm km\,s^{-1}}$ ($\beta\approx0.100$) and matter at $\sim 40000\,{\rm km\,s^{-1}}$ 
($\beta\approx 0.133$) may be the highest ever needed \citep{jeffery1992}.   
For such velocities, relativistic effects are small since they mostly go as $\beta^{2}$.
Using our fiducial value $v_{e}=2894.7\,{\rm km\,s^{-1}}$, 
the reduced velocity corresponding to $30000\,{\rm km\,s^{-1}}$ is $\sim 10$.
From equation~(\ref{eq-interior-mass-fraction}), $f(10)\approx 0.997$.
Thus, in modeling with exponential density profiles, typically much less than $1\,$\% of the ejecta
will be put at velocities where relativistic effects are important. 
Therefore, typical relativistic effects are going to be negligible when using exponential density profiles.

       As a semi-necessary adjunct to assuming an exponential density profile, 
we also assume that the ejecta can be approximated as spherically symmetric.
This is a reasonable simplifying assumption, but is not necessarily completely valid.
Because of their homogeneity, normal SNe~Ia are probably quite spherically symmetric:  even identical SNe~Ia 
would look significantly different from different viewing directions if they were significantly aspherical.
There is, however, some evidence from spectropolarimetry for departures from spherical symmetry for both
normal SNe~Ia \citep{wang2003} and peculiar SNe~Ia \citep{howell2001}.
Thus, a~priori, one cannot rule out significant asymmetry for any particular SN~Ia especially if it is unprecedented
like SN~2003fg.
Moreover, if the progenitor of SN~2003fg was indeed rotating and this accounts for its probable super-Chandrasekhar mass,
then its progenitor could have been quite aspherical. 
The ratio of polar to equatorial radii for YL's rotating WD models fall as low as $\sim 0.3$ 
for masses of $\sim 2\,M_{\odot}$ (see YL's Tables~1 and~2).
The explosion of such asymmetric WDs may, however, reduce the asymmetry of the matter considerably (YL).  
For the present, assuming spherical symmetry for the ejecta is justified by simplicity and the lack of
any evidence that SN~2003fg shows any asymmetry.

       The third component of the SSC~model is to assume that a well-defined
IPE core exists with a boundary that we can approximate as absolutely sharp and 
that we can make the approximation that all explosion-synthesized IPE matter is within the IPE core and no
other elements are. 
This approximation is reasonable as discussed in \S~\ref{section-IPE-core-boundary}.

       Following a similar path to H2006, we take ejecta mass $M$, IPE-core mass $M_{\rm IPE\mathhyphen core}$,
a new parameter $h$, and $\rho_{\rm central,WD}$ as 4 independent parameters.
The new parameter $h$ is given by
\begin{equation}
         h={f_{\rm IME}\over f_{\rm CO}+f_{\rm IME} } \,\, .
\end{equation}
We prefer $h$ as an independent parameter rather than $f_{\rm CO}$ adopted by H2006 or $f_{\rm IME}$ which
could also be used.
One can obtain $f_{\rm CO}$ and $f_{\rm IME}$ from $h$ and $f_{\rm IPE}$
(which is obtained from equation~(\ref{eq-fipe-fime}) (\S~\ref{section-mass}) using $M$ and $M_{\rm IPE\mathhyphen core}$)
using the formulae
\begin{equation}
         f_{\rm CO}=(1-h) (1-f_{\rm IPE}) \qquad\hbox{and}\qquad
         f_{\rm IME}=h(1-f_{\rm IPE})  \,\, .
\label{eq-fco-fime}
\end{equation}
Using $h$ as an independent parameter makes simpler the use of IPE-core velocity $v_{\rm IPE\mathhyphen core}$ as an independent
parameter in place of $M$.
We want this simplification when we solve for an unknown $M$ given $M_{\rm IPE\mathhyphen core}$, 
$v_{\rm IPE\mathhyphen core}$, $h$, and $\rho_{\rm central,WD}$.
We find such solutions in \S~\ref{section-comparison-sn2003fg} and show how the solutions
are done and why $h$ preferred to $f_{\rm CO}$ and $f_{\rm IME}$ as an independent parameter in Appendix~\ref{ap-SSC-mass}. 

    For convenient reference and making use
of equations~(\ref{eq-fipe-fime}) (\S~\ref{section-mass}) and~(\ref{eq-fco-fime}), 
we rewrite the kinetic energy equation~(\ref{eq-energy-a}) (\S~\ref{section-mass}) 
in terms of parameter $h$:
\begin{equation}
         E=E_{\rm IPE}M[f_{\rm IPE}+Rh(1-f_{\rm IPE})] - E_{\rm bind}(\rho_{\rm central,WD},M)  \,\, . 
\label{eq-energy-b}
\end{equation}
Substituting values for $M$, $f_{\rm IPE}$, $h$, and $\rho_{\rm central,WD}$ 
into equation~(\ref{eq-energy-b}), we get the kinetic energy $E$ and 
then substituting values for $M$ and $E$ into equation~(\ref{eq-efolding}) gives the $e$-folding velocity $v_{e}$.

      Now to determine $v_{\rm IPE\mathhyphen core}$, we need the reduced IPE-core velocity $x_{\rm IPE\mathhyphen core}$ which we obtain
by solving for $x_{\rm IPE\mathhyphen core}$ from equation~(\ref{eq-interior-mass-fraction}) with $f$ set to $f_{\rm IPE}$.
Unfortunately, there is no general analytic solution for $x$ given $f$ for equation~(\ref{eq-interior-mass-fraction}).
In the special cases of $f=0$ and $f=1$, the solutions are, respectively,  $x=0$ and $x=\infty$.
Otherwise we can solve equation~(\ref{eq-interior-mass-fraction}) by the Newton-Raphson method \citep[e.g.,][p.~355ff]{press1992}
for $x$ given $f$.
The Newton-Raphson method is guaranteed to converge in this case since $f$ is monotonically increasing with $x$ and
only has a stationary point at infinity provided one prevents the Newton-Raphson corrections
from causing overshooting of the diminishing allowed range for the final solution.    
The derivative of $f$ (which is needed for the Newton-Raphson method solution) is
\begin{equation}
       {df\over dx}={x^{2}\over2}\exp(-x) \,\, .
\end{equation}
A good initial value for the Newton-Raphson method iteration for $f<1$ is the 3rd-order-in-small-$y$ solution for $x$
given by 
\begin{equation}
       x_{\rm 3rd}=y+{1\over4}y^{2}+{7\over80}y^{3}  \,\, , 
\label{eq-x-3rd}
\end{equation}
where $y=(6f)^{1/3}$.
This solution for $x$ is about $10\,$\% accurate for $f\lesssim 0.33$:  it improves in accuracy as $y\to0$,
of course.
Either the 2nd-order-in-small-$y$ or 1st-order-in-small-$y$ solutions for $x$ are almost as good initial
values for the iteration for any $f<1$ as the 3rd-order-in-small-$y$ solution for $x$.
As an alternative to the Newton-Raphson method, one can solve for $x$ using the iteration equation method using
an iteration equation obtained from rearranging equation~(\ref{eq-interior-mass-fraction}).
The iteration equation is
\begin{equation}
       x_{\rm out}=\ln\left[{1+x_{\rm in}+(1/2)x_{\rm in}^{2}\over 1-f}\right] \,\, , 
\end{equation}
where $x_{\rm in}$ is an input $x$ value and $x_{\rm out}$ is an output $x$ value.
The iteration is guaranteed to converge since the derivative $|dx_{\rm out}/dx_{\rm in}|\leq1$ with
the equality holding only for $x_{\rm in}=0$ which is a point that can be avoided since we already know
that $x=0$ for $f=0$. 
In practice, the iteration equation solution to reasonable convergence takes about $50\,$\% longer in CPU
time than the Newton-Raphson method.   
One can obtain a reasonable approximate analytic solution for $x$, $x_{\rm approx}$, by substituting $x_{\rm 3rd}$ 
from equation~(\ref{eq-x-3rd})
into the iteration equation.
This approximate analytic solution is given by
\begin{equation}
       x_{\rm approx}=\ln\left[{1+x_{\rm 3rd}+(1/2)x_{\rm 3rd}^{2}\over 1-f}\right] \,\,  .
\label{eq-x-approx}
\end{equation}
The approximate analytic solution is always an underestimate with a maximum relative error 
of $\sim 19\,$\% for $f\approx0.995$:  the
relative error decreases with $f$ increasing or decreasing from $f\approx 0.995$ and it goes to zero in
the limits $f\to0$ and $f\to1$.
More elaborate analytic formulae can give better accuracy for $x$ than
equation~(\ref{eq-x-approx}):  e.g., substituting $x_{\rm approx}$ itself into 
equation~(\ref{eq-x-approx}) instead of $x_{\rm 3rd}$ gives a significantly more accurate approximate analytic
formula for $x$.
By whatever means, having obtained an accurate $x_{\rm IPE\mathhyphen core}$, we
obtain the IPE-core velocity from 
\begin{equation}
     v_{\rm IPE\mathhyphen core}=x_{\rm IPE\mathhyphen core}v_{e} \,\, .
\end{equation}

       H2006 in their analysis (see \S~\ref{section-mass})
compared 40-days-past-$B$-maximum-light Si~II~$\lambda6355$ line velocities to $v_{\rm kin}$.
As we argued in \S~\ref{section-mass} and \S~\ref{section-IPE-core-boundary}, the
long-past-$B$-maximum-light Si~II~$\lambda6355$ line velocity is a reasonable approximation
to the IPE-core velocity.
As H2006 acknowledged, their comparison was ambiguous because there is no necessary theoretical
connection between the long-past-$B$-maximum-light Si~II~$\lambda6355$ line velocity and $v_{\rm kin}$.
The situation is further clarified by the SSC~model, where it is clear that
there is no necessity for $v_{\rm IPE\mathhyphen core}$ to equal $v_{\rm kin}$ even approximately although it could do so. 
If we set $v_{\rm IPE\mathhyphen core}$ to $v_{\rm kin}$ (i.e., $x_{\rm IPE\mathhyphen core}=\sqrt{12}$), 
then $f_{\rm IPE}$ would be fixed at
$0.6724619703381\approx 2/3$ as we see from equation~(\ref{eq-interior-mass-fraction}).
However, in the SSC~model we are free to vary $f_{\rm IPE}$ from 0 to 1 provided that $E\geq0$ which is
necessary for the WD to explode.

      We illustrate the behavior of the SSC model in Figures~\ref{f1} and~\ref{f2}
by plotting $v_{\rm IPE\mathhyphen core}$ as a function of $M_{\rm IPE\mathhyphen core}$ for
representative values of the 3 other independent parameters (i.e., mass $M$, $h$, and $\rho_{\rm central,WD}$)
which are held constant for each curve.
(The data points on the figures are discussed in 
\S\S~\ref{section-comparison-sample} and~\ref{section-comparison-sn2003fg}.)
Recall from \S~\ref{section-mass} that
the $\rho_{\rm central,WD}$ range $\sim 2.5\times10^{9}$---$10^{10}\,{\rm g\,cm^{-3}}$ is believed to be
the allowed range for WD explosions and that our fiducial $\rho_{\rm central,WD}=2.9\times10^{9}\,{\rm g\,cm^{-3}}$.
To explore the possible $\rho_{\rm central,WD}$ range for SN~Ia progenitors,
we used the fiducial $\rho_{\rm central,WD}=2.9\times10^{9}\,{\rm g\,cm^{-3}}$ 
(which is not far from the low end of the range)
for the Figure~\ref{f1} calculations and
$\rho_{\rm central,WD}=10^{10}\,{\rm g\,cm^{-3}}$ for the Figure~\ref{f2} calculations.
Most explosion modelers find $\rho_{\rm central,WD}$ much closer to the low end of the allowed range
for plausible explosions \citep[e.g.,][]{nomoto1984,thielemann1986,khokhlov1993,woosley1994,
hoeflich1996,hoeflich1998}:   
thus $\rho_{\rm central,WD}=10^{10}\,{\rm g\,cm^{-3}}$ is actually an extreme upper
limit and a WD that evades an explosion and gets to this limit is probably on the verge of collapse 
to a neutron star (\citealt{nomoto1991}; YL). 
The binding energy increases with central density (see YL's Fig.~9), and so all other things being equal,
increasing density leads to lower $v_{\rm IPE\mathhyphen core}$.
This is why the curves in Figure~2 are generally lower than their counterparts in Figure~1.

    The solid curves (which form cornucopias) are for $h=1$ which means $f_{\rm IME}=1-f_{\rm IPE}$,
$f_{\rm CO}=0$, and there are only IMEs above the IPE core.
The mass in solar mass units for each solid curve is given at the upper right end of the curve:
the masses run from $1.388\,M_{\odot}$ for Figure~\ref{f1} and $1.414\,M_{\odot}$ for Figure~\ref{f2}
to $3\,M_{\odot}$ in increments of $0.1\,M_{\odot}$ for $M\geq 1.5\,M_{\odot}$.
The masses $1.388\,M_{\odot}$ and $1.414\,M_{\odot}$ are 
the masses for non-rotating WDs from YL's equation~(22) (see also eq.~(\ref{eq-mass-nr}) in Appendix~\ref{ap-YL-formula})
for the input $\rho_{\rm central,WD}$ values and we just label their curves by $M_{\rm NR}$ on the figures.
The non-rotating WD masses are the lower bounds on the masses for the input $\rho_{\rm central,WD}$ values
and are near the physical Chandrasekhar mass $1.38\,M_{\odot}$ (see Appendix~\ref{ap-chan-mass}).
As mentioned in \S~\ref{section-mass}, the non-rotating WD masses can slightly exceed the physical
Chandrasekhar mass $1.38\,M_{\odot}$.
This is because YL use an approximate equation of state for their modeling that gives a Chandrasekhar
mass of $1.436\,M_{\odot}$ (see Appendices~\ref{ap-chan-mass} and~\ref{ap-YL-formula}).
This inconsistency from the physical Chandrasekhar mass is small in comparison to the mass variations we
are considering.
Actually all the masses we discuss probably have uncertainties of a few percent for the given
input parameters because they are based on YL's choice of equation of state:  
they are probably mostly slightly larger than the input parameters would
imply with a more accurate equation of state (see Appendix~\ref{ap-chan-mass}).
For clarity, we call the $1.388\,M_{\odot}$ and $1.414\,M_{\odot}$ mass curves the $M_{\rm NR}$ mass curves.

      The general behavior of the solid curves can be explained as follows.
In regard to varying mass $M$ for a given IPE-core mass, the smaller $M$ is, the higher $v_{\rm IPE\mathhyphen core}$ 
tends to be since the
WD binding energy decreases with decreasing $M$, there is less mass to accelerate, $v_{\rm IPE\mathhyphen core}$
moves to a higher mass fraction in the ejecta,
and the nuclear burning energy decreases rather weakly only through decreasing amount of burning to IMEs.
This explains why as curve mass decreases, the curves move upward.
In overall curve appearance, the curves move toward the upper left as mass decreases.
In regard to varying $M_{\rm IPE\mathhyphen core}$ for fixed $M$, as $M_{\rm IPE\mathhyphen core}$ increases, 
$v_{\rm IPE\mathhyphen core}$ increases monotonically.
There are two causes for this.
First, as one moves outward in mass fraction with everything fixed, ejecta velocity will increase since
outer matter must move faster than inner matter.
Second if as one moves outward in mass fraction and the IPE-core boundary moves with you, there is more kinetic
energy than otherwise to drive the outer layers.
Both causes help $v_{\rm IPE\mathhyphen core}$ to increase monotonically with $M_{\rm IPE\mathhyphen core}$.
Note that the solid curves all begin at $M_{\rm IPE\mathhyphen core}=0$ since burning all the
non-IPE-core matter to IMEs always ensures adequate energy for a WD explosion.
The solid curves also begin at $v_{\rm IPE\mathhyphen core}=0$ since that is the limiting IPE-core velocity
when there is no IPE core.

     The dotted curves are for $h=0$ which means $f_{\rm IME}=0$,
$f_{\rm CO}=1-f_{\rm IPE}$, and there is only unburnt CO above the IPE core.
As mentioned in \S~\ref{section-introduction},   
the spectra of all SNe~Ia, including SN~2003fg, are
dominated by IME lines at least in the early-post-$B$-maximum phase, 
and so we know that there must be abundant IMEs in the outer layers of all 
SN~Ia ejecta.
Partially successful explosion models also yield abundant IMEs:  for example, Model~W7 gives $f_{\rm IME}=0.281$
and $h=0.798$ \citep{thielemann1986};  
example delayed-detonation models give $h\gtrsim 0.95$ \citep{khokhlov1993,hoeflich1998}.  
Thus, $0$ must be regarded an extreme lower limit on $h$.
However, as discussed in \S~\ref{section-mass}, SN~2003fg may have had more unburnt CO than 
normal SNe~Ia, and so $h$ may be significantly less than 1 for SN~2003fg. 
Thus, the $h=0$ curves are relevant to this study.

     The masses for the dotted curves are the same as for the solid curves
and dotted curves are ordered in the same way as the solid curves  by mass:  the dotted
curves are labeled by mass at their low IPE-core mass end.
On Figure~\ref{f2} the dotted curves for masses greater than $2.8\,M_{\odot}$ are off the figure.
Since there is no burning of IMEs for the dotted curves, they represent
models with less energy than the counterpart solid curves.
Thus, $v_{\rm IPE\mathhyphen core}$ values for the dotted curves at a given $M_{\rm IPE\mathhyphen core}$ 
are always less than that for the
counterpart solid curves at the same $M_{\rm IPE\mathhyphen core}$.
The dotted curves all begin at zero IPE-core velocity and at non-zero IPE-core mass.
In the case of the dotted curves, the IPE core must reach a sufficient size before there
is enough nuclear burning energy to unbind the WD:  recall there is no burning of matter outside of the IPE core
in the case of these curves.
The behaviors of the dotted curves as $M$ is varied for fixed $M_{\rm IPE\mathhyphen core}$ and as 
$M_{\rm IPE\mathhyphen core}$ is
varied for fixed $M$ have essentially the same explanation as for the solid curves. 
 
     As $M_{\rm IPE\mathhyphen core}$ grows for a constant $M$, the relative amounts of IMEs and CO become
less important for the energetics, and thus the dotted curves converge toward their counterpart solid curves as
$M_{\rm IPE\mathhyphen core}$ grows large. 
For the counterpart dotted and solid curves for larger $M$, close convergence 
occurs off the top or right edges of the figures.
Final convergence, of course, happens in the limit of $M_{\rm IPE\mathhyphen core}\to M$
which causes $v_{\rm IPE\mathhyphen core}$ to go to the unphysical limit of infinity.
Non-relativistic physics fails long before this velocity limit is reached.
Since the figures only show up to $13000\,{\rm km\,s^{-1}}$ (which is $\beta\approx0.0434$),
the region where non-relativistic physics fails occurs well off the figures for all the curves.

     We need to emphasize that YL have only verified their binding energy formula for 
the mass range $\sim 1.4$--$2.05\,M_{\odot}$, and thus we have made a large extrapolation
of the usage of binding energy formula in plotting curves for masses up to $3\,M_{\odot}$.

\section{COMPARING THE SSC MODEL AND OBSERVATIONS OF THE LOW-$z$ SAMPLE OF SNe Ia\label{section-comparison-sample}}

        The circle points on Figures~\ref{f1} and~\ref{f2} are for the low-$z$ sample of SNe~Ia from
\citet{benetti2005} which was discussed in \S~\ref{section-mass}.
The $+$ points are discussed below, and the letters N, O, X, and Y are for SN~2003fg 
and are discussed in \S~\ref{section-comparison-sn2003fg}. 
The IPE-core masses for the circle points were determined using equation~(\ref{eq-IPE-core-mass}) with the 
$L_{\rm bol}/\dot E_{\rm ^{56}Ni}(t_{\rm bol})$ 
values of H2006, $\alpha=1.2$ (as adopted by H2006 and which
we take to be fiducial for this paper), and $g=0.7$ (which, following H2006, we argued
was a good representative fiducial value in \S~\ref{section-mass}).
The IPE-core masses are in the range $\sim 0.04$--$0.86\,M_{\odot}$.
The velocities are the Si~II~$\lambda6355$ line velocities from 35 days past $B$ maximum derived
using the least-squares line fits of \citet{benetti2005}.
The velocities are in the range $\sim 5900$--$11200\,{\rm km\,s^{-1}}$. 
As mentioned in \S~\ref{section-mass}, the uncertainties of 
$L_{\rm bol}/\dot E_{\rm ^{56}Ni}(t_{\rm bol})$ 
for the low-$z$ sample are in general
greater than for SN~2003fg because some of these SNe~Ia are not well out in the Hubble flow.  
We will neglect this complication in our analysis for simplicity and since we are primarily just 
interested in overall behavior of the low-$z$ sample when analyzed using the curves on Figures~\ref{f1} and~\ref{f2}. 

     We prefer 35 days past $B$~maximum to H2006's 40 days past $B$~maximum in calculating the
Si~II~$\lambda6355$ line velocities to be used in comparison to the SSC~model curves.
The reasoning is that after about 35 days past $B$~maximum the derived values in most cases would be extrapolations of
uncertain quality of the Si~II~$\lambda6355$ line velocity curves \citep[Fig.~1]{benetti2005}. 
Also some of the Si~II~$\lambda6355$ line velocity curves seem to be plateauing at about
35 days past $B$~maximum \citep[Fig.~1]{benetti2005} which suggests that the Si~II~$\lambda6355$ line
formation layer has stopped receding into the ejecta
because it has receded to the boundary of IPE core and there is no significant silicon below that layer.
Thus, the 35-days-past-$B$--maximum values may in many cases best approximate the IPE-core velocities
and this is what we hope the long-past-$B$-maximum Si~II~$\lambda6355$ line velocities
will approximate. 

        The Si~II~$\lambda6355$ line velocity of SN~1991bg may still be decreasing at
35 days past $B$~maximum which is about as far as the data of \citet{benetti2005} go, and so SN~1991bg's 
35-day-past-$B$-maximum Si~II~$\lambda6355$ line velocity is only an upper limit on the IPE-core velocity.
From modeling, \citet{mazzali1998} found that the IPE-core velocity of SN~1991bg was $\sim 2000\,{\rm km\,s^{-1}}$.
SN~1991bg is one of the 5 SNe~Ia designated as faint by \citet{benetti2005} in the low-$z$ sample.
These faint SNe~Ia have the lowest IPE-core masses, and so give the leftmost circle data points on
Figures~\ref{f1} and~\ref{f2}.
In order of increasing IPE-core mass on Figures~\ref{f1} and~\ref{f2}, the faint SNe~Ia are:   SN~1999by, SN~1991bg,
SN~1997cn, SN~1986g, and SN~1993H.
The situation for the Si~II~$\lambda6355$ line velocities of the other faint SNe~Ia
are not as clear as for SN~1991bg, but their
day-35-past-$B$-maximum Si~II~$\lambda6355$ line velocities are more likely to
be only upper bounds on the IPE-core velocities than is the case for the
other SNe~Ia in the low-$z$ sample.

        The circle points are rather dispersed and do not lie along the $M_{\rm NR}$ mass curves.
Some are to the right of the $M_{\rm NR}$ mass curves which suggests the possibility that they
are for slightly super-Chandrasekhar SNe~Ia.
But this is very uncertain.
The circle points to the left of the $M_{\rm NR}$ mass curves are not to be taken
for (significantly) sub-Chandrasekhar SNe~Ia.
A mechanism for sub-Chandrasekhar SNe~Ia has been proposed \citep{livne1990,livne1991}, but
this mechanism has been ruled out.
Explosion models produced using the mechanism led to synthetic lightcurves \citep{hoeflich1996} 
and spectra \citep{nugent1997} that failed to match observations:  among other things the synthetic
lightcurves and spectra were too blue.  
Thus, the circle points to the left of the $M_{\rm NR}$ mass curves are all likely to be from
near-Chandrasekhar (or less likely somewhat super-Chandrasekhar) SNe~Ia.
The main explanation for the deviations of the circle points from the $M_{\rm NR}$ mass curves
is likely that $\alpha$ is not $1.2$ in general.
The faint group of SNe~Ia may also deviate significantly from the $M_{\rm NR}$ mass curves because their
day-35-past-$B$-maximum Si~II~$\lambda6355$ line velocities are not close upper bounds on
their IPE-core velocities. 
Of course, the SSC~model may also be somewhat inadequate for real SNe~Ia and/or the $g$ factor 
may vary significantly from the fiducial $0.7$ value. 

       The reason for believing it is varying $\alpha$ that leads to the dispersion
of the circle points from the $M_{\rm NR}$ mass curves is, as discussed in \S~\ref{section-mass}, that there is reason to believe
that $\alpha$ can have a large range:   recall that the modeling results of \citet{hoeflich1996}
suggest a range of $0.62$--$1.46$ for $\alpha$.
We can derive a semi-empirical range for $\alpha$, by varying $\alpha$ for the low-$z$ sample 
such that all the points lie on the $M_{\rm NR}$ mass curves.
The range is semi-empirical because we are relying on observations and on the SSC~model
in the non-rotating mass limit.
The range is based on the assumption that most SNe~Ia have nearly the Chandrasekhar mass (which
is approximated by $M_{\rm NR}$ in the SSC~model). 
We will only do the variation for the $M_{\rm NR}$ mass curves for $h=1$ since this
is likely to be closer to the truth for normal SNe~Ia than curves for $h=0$ (see \S~\ref{section-SSC}).  
The $+$ points on Figures~\ref{f1} and~\ref{f2} are the result of varying $\alpha$.
Note that $\alpha<1.2$ moves points to larger IPE-core mass and
$\alpha>1.2$ moves points to smaller IPE-core mass.
The range for $\alpha$ is $0.13$--$1.72$ for Figure~1 and  $0.13$--$1.67$ for Figure~2.
If we exclude the five faint SNe~Ia for which the day-35-past-$B$-maximum Si~II~$\lambda6355$ line velocities
are more likely to be only upper bounds on the IPE-core velocities than for the other SNe~Ia,
we find the $\alpha$ ranges to be $0.68$--$1.72$ for Figure~1 and  $0.65$--$1.67$ for Figure~2.
Given the assumptions that the SSC~model is valid, the non-faint SNe~Ia in the low-$z$ sample have
day-35-past-$B$-maximum Si~II~$\lambda6355$ line velocities
are nearly the IPE-core velocities, the $g$ factor is $0.7$, $h\approx 1$,
and that none of the low-$z$ sample is significantly
super-Chandrasekhar (nor sub-Chandrasekhar), then the semi-empirical $\alpha$ range can be taken as $\sim 0.6$--$1.7$. 
This $\alpha$ range is plausible given the results of \citet{hoeflich1996}, but it is not definitive,
of course. 
      If we used the $h=0$ curves instead of the $h=1$ curves,
the semi-empirical range for $\alpha$ would shift to lower values.

\section{COMPARING THE SSC MODEL TO THE SN 2003fg OBSERVATIONS\label{section-comparison-sn2003fg}}

\begin{figure}
\includegraphics[height=.7\textheight,angle=-90]{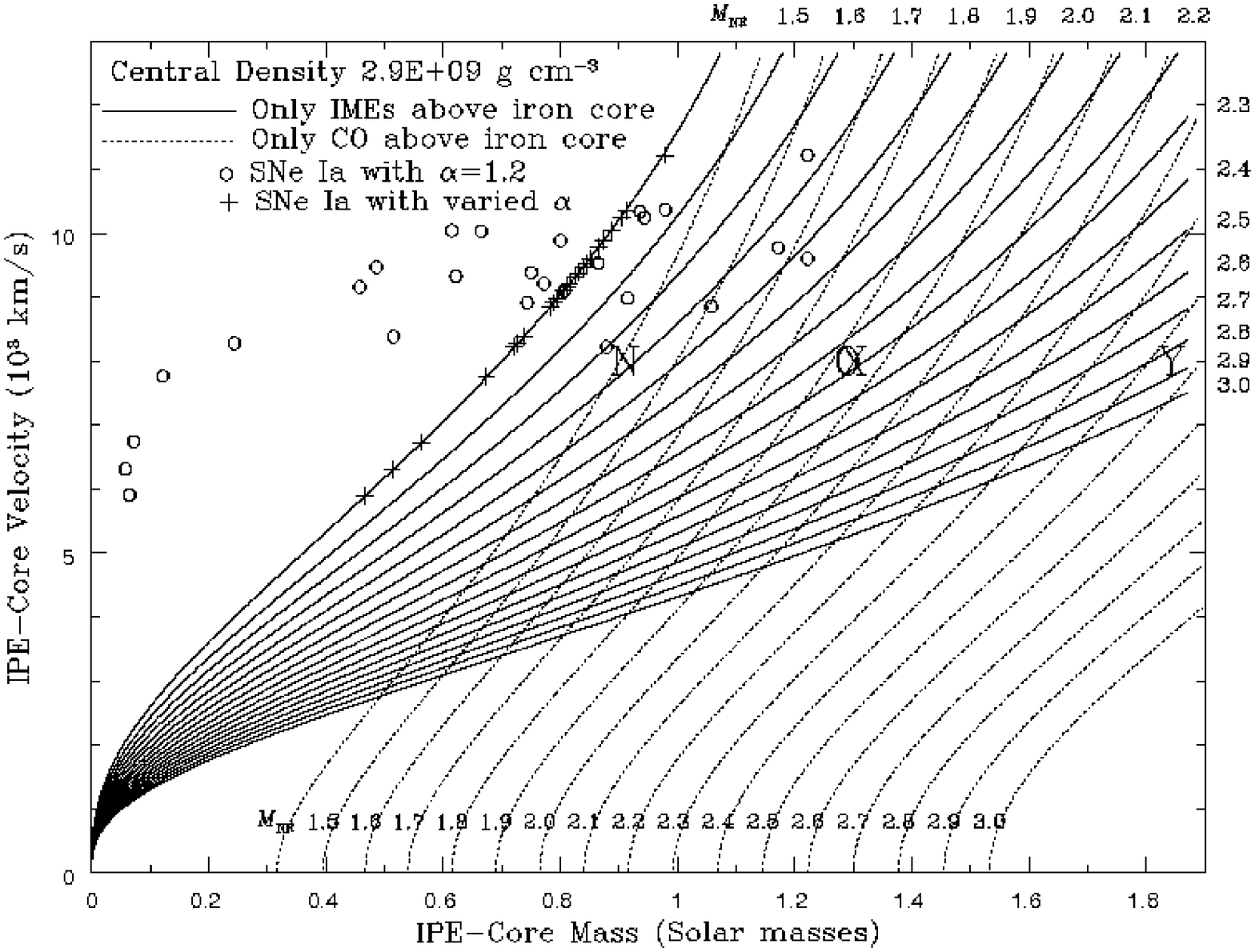}
\caption{
Curves of constant SN~Ia mass from the SSC model for super-Chandrasekhar-mass
SNe~Ia on a plot of IPE-core velocity versus IPE-core mass.
There are two families of curves:  the solid curves are for only IMEs above
the IPE core ($h=1$) and the dotted curves are for only CO above the IPE core ($h=0$).
This plot is calculated for WD progenitors with central density $2.9\times10^{9}\,{\rm g\,cm^{-3}}$.
The plot and the data points are explained at length in the text
(\S\S~\ref{section-SSC}, \ref{section-comparison-sample}, and~\ref{section-comparison-sn2003fg}).
\label{f1}}
\end{figure}

\begin{figure}
\includegraphics[height=.7\textheight,angle=-90]{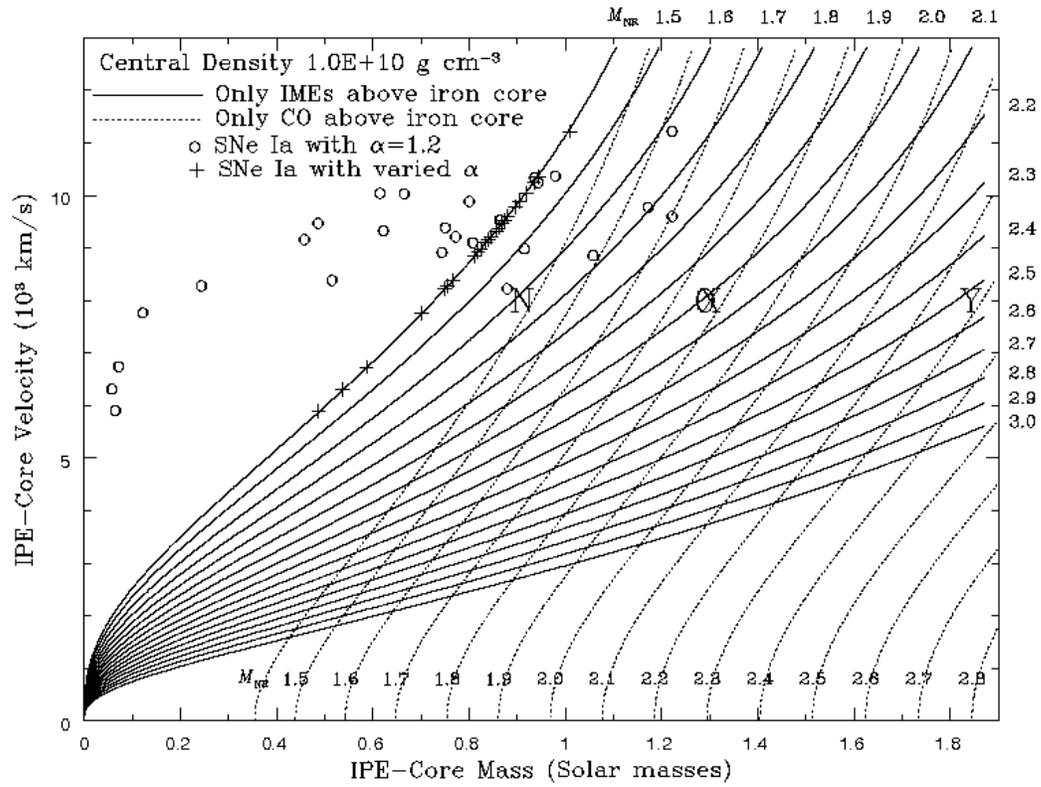}
\caption{
The same as Figure~\ref{f1}, except that the
plot is calculated for WD progenitors with central density $10^{10}\,{\rm g\,cm^{-3}}$.
\label{f2}}
\end{figure}

        We now turn our attention to comparing the SSC~model to the SN~2003fg observations. 
In the case of SN~2003fg, we cannot adequately determine the 
day-35-past-$B$-maximum Si~II~$\lambda6355$ line velocity given only the near-$B$-maximum
Si~II~$\lambda6355$ line velocity of $8000\pm 500\,{\rm km\,s^{-1}}$.
Therefore, we will only adopt $8000\,{\rm km\,s^{-1}}$ as an upper bound on the
IPE-core velocity.
The points N, O, X (partially overlapping with O), and Y on Figures~\ref{f1} and~\ref{f2} are
all SN~2003fg IPE-core velocity upper bound points of $8000\,{\rm km\,s^{-1}}$.
As Figures~\ref{f1} and~\ref{f2} show, since we can only set an upper bound on the SN~2003fg
IPE-core velocity, we can only set a lower bound on the SN~2003fg mass.

    Unfortunately in trying to set such a lower bound on SN~2003fg mass, 
there is no secure way to estimate the IPE-core mass $M_{\rm IPE\mathhyphen core}$ (which is a SSC~model parameter) 
primarily because of the dispersion in $\alpha$
and secondarily because we can only estimate the $g$ factor.
Recall $M_{\rm IPE\mathhyphen core}$ is calculated from 
equation~(\ref{eq-IPE-core-mass}) in \S~\ref{section-mass} using $g$ and $\alpha$
and H2006's 
$L_{\rm bol}/\dot E_{\rm ^{56}Ni}(t_{\rm bol})$
value $1.55\,M_{\odot}$ which we
assume to have negligible uncertainty. 
The values of the SSC~model parameters
$\rho_{\rm central,WD}$ and $h$ are also not fully constrained.
Therefore, we calculate lower bound masses varying the four parameters
$\rho_{\rm central,WD}$, $g$, $\alpha$, and $h$ from what we believe to
good representative fiducial values to what we believe to be extreme values which
tend to minimize the lower bound masses.
(Recall, they will be lower bound masses because $8000\,{\rm km\,s^{-1}}$ which we use as the input
value for $v_{\rm IPE\mathhyphen core}$  is only an upper
bound on $v_{\rm IPE\mathhyphen core}$.)
The fiducial values are 
$\rho_{\rm central,WD}=2.9\times10^{9}\,{\rm g\,cm^{-3}}$ 
(which we adopted as fiducial in \S~\ref{section-mass} following \citet{woosley2006}),  
$g=0.7$ (which is certainly good for some SN~Ia models:  see \S~\ref{section-mass}),
$\alpha=1.2$ (which is good in the sense that it is not extreme:  see \S\S~\ref{section-mass} and~\ref{section-comparison-sample})
and
$h=1$ (which is good for normal SNe~Ia given that their spectra up until after $B$~maximum are dominated by IME lines 
and given that model~W7 has $h=0.798$ \citep{thielemann1986} and some example delayed-detonation models
have $h\gtrsim 0.95$ \citep[e.g.,][]{khokhlov1993,hoeflich1998}).
The extreme parameter values we consider are
$\rho_{\rm central,WD}=10^{10}\,{\rm g\,cm^{-3}}$
(which is extreme based on the fact the WD is on the verge of collapse rather than explosion
for this value:  \citet{nomoto1991};  YL),
$g=1$ (which is an absolute upper bound:  see \S~\ref{section-mass}),
$\alpha=1.7$ (which is the upper end of the semi-empirical range for $\alpha$ that we derived 
\S~\ref{section-comparison-sample} and
is larger than the largest $\alpha$ value of \citet{hoeflich1996} (see \S~\ref{section-mass})),
and
$h=0$
(which is extreme for normal SNe~Ia given that spectra until after $B$-maximum are dominated by IME lines and for SN~2003fg given 
its early-post-$B$-maximum spectrum is fairly normal, and so dominated by IME lines).

     Given two values of each of 
$\rho_{\rm central,WD}$ (i.e., $2.9\times10^{9}\,{\rm g\,cm^{-3}}$ and $10^{10}\,{\rm g\,cm^{-3}}$), 
$g$ (i.e., $0.7$ and $1$), 
$\alpha$ (i.e., $1.2$ and $1.7$), 
and
$h$ (i.e., $1$ and $0$),
there are 16 possibilities which we represent
by the 8 SN~2003fg points on the Figures~\ref{f1} and~\ref{f2}:  
N ($g=1$, $\alpha=1.7$),
O ($g=1$, $\alpha=1.2$), 
X ($g=0.7$, $\alpha=1.7$), 
and 
Y ($g=0.7$, $\alpha=1.2$).

     The input parameter values for all the SN~2003fg points and the lower bound masses we derive from
them are given in Table~\ref{table1}. 
Table~\ref{table1} also gives other output parameter values for the input parameter value sets:  
i.e., $f_{\rm CO}$, $f_{\rm IME}$, $f_{\rm IPE}$, 
$f_{\rm ^{56}Ni}$ 
(which is the mass fraction of $^{56}$Ni),
$v_{e}$, and $E$. 
The lower bound masses are actually obtained by solving SSC~model using
$M_{\rm IPE\mathhyphen core}$, $v_{\rm IPE\mathhyphen core}$, 
$\rho_{\rm central,WD}$,
and $h$
as 4 independent parameters and the procedure given in Appendix~\ref{ap-SSC-mass}. 
The last-line parameter values with $v_{\rm IPE\mathhyphen core}=8000\,{\rm km\,s^{-1}}$ give no solution for the mass 
since even a non-rotating WD for the
given input parameters is too massive to yield this $v_{\rm IPE\mathhyphen core}$ value.
So we set $v_{\rm IPE\mathhyphen core}=7639.5\,{\rm km\,s^{-1}}$ which is the largest $v_{\rm IPE\mathhyphen core}$ for which
a solution exists given the other input parameter values.
The output mass is just the mass of a non-rotating WD.
The output masses of Table~\ref{table1} can also be obtained, of course,
by interpolation from Figures~\ref{f1} and~\ref{f2}.

    We regard all the output masses in Table~\ref{table1} as lower bounds on the actual mass
since, as emphasized above, $8000\,{\rm km\,s^{-1}}$ is an upper bound on the IPE-core velocity and
$7402.947\,{\rm km\,s^{-1}}$ is either an upper bound or at least not far removed from an upper bound.

    The lower bound masses in Table~\ref{table1}
are rated by the number of extreme parameter values (EPVs) used to calculate them:  the
lower the number, the more probable we think the lower bound mass.
The most probable set of parameters is for point~Y on Figure~\ref{f1} (0 EPVs)
and the least probable is for point~N on Figure~\ref{f2} (4 EPVs) for which in fact
there is actually no $M$ solution for $v_{\rm IPE\mathhyphen core}=8000\,{\rm km\,s^{-1}}$ as noted above.
The least probable set of parameters for a point on Figure~\ref{f1} are for point~N (3 EPVs). 

    We see that the least probable lower bound masses for each figure (which are those associated with the N points)
are approximately the Chandrasekhar mass. 
Excluding these lower bound masses gives a lowest lower bound mass of $1.62\,M_{\odot}$.  
Excluding lower bound masses with two or more EPVs leaves 5 lower bound masses:  the lowest of these is $2.16\,M_{\odot}$.
The formally most probable lower bound mass (associated with point~Y on Fig.~\ref{f1}) according to our rating
system is the highest one, $2.84\,M_{\odot}$.
This lower bound mass is close to the mass estimate of $\sim 3\,M_{\odot}$ for SN~2003fg that
we obtained from the simple-scaling argument (\S~\ref{section-mass}). 

     From the above values we conclude that it is very probable that SN~2003fg is super-Chandrasekhar.
The only lower bound masses that are approximately the Chandrasekhar mass are for extreme and improbable sets of the parameters.
The more probable sets give lower bound masses that are super-Chandrasekhar and the 5 most probable sets
give lower bound masses that are $\gtrsim 2\,M_{\odot}$.
Given that YL's binding energy formula is only verified for $M$ in the range $\sim 1.4$--$2.05\,M_{\odot}$, 
a SN~2003fg mass $\gtrsim 2\,M_{\odot}$ is as much as one can say 
based YL's binding energy formula for rotating super-Chandrasekhar-mass
WDs (Appendix~\ref{ap-YL-formula}) and the SSC~model (\S~\ref{section-SSC}).
Given that the analysis of \citet{langer2000} of accretion histories of WDs in binary systems limits rotating  
WD mass to $\lesssim 2\,M_{\odot}$, our analysis leads to the final conclusion that the mass of
SN~2003fg is probably $\sim 2\,M_{\odot}$.

      The limitation on mass from \citet{langer2000}, has implications for the possible parameter values for 
$g$ and $\alpha$.
In order to have the SN~2003fg mass $\sim 2\,M_{\odot}$, the product of $g\alpha$ would have to be more than
the fiducial value of $0.7\times1.2=0.84$.
Demanding $M=2\,M_{\odot}$, solving for $M_{\rm IPE\mathhyphen core}$ for this $M$ using 
the $v_{\rm IPE\mathhyphen core}=8000\,{\rm km\,s^{-1}}$ and
the $\rho_{\rm central,WD}$ and $h$ values of this section and Newton-Raphson method procedure of
Appendix~\ref{ap-SSC-IPE-core-mass}, and using H2006's value for
$L_{\rm bol}/\dot E_{\rm ^{56}Ni}(t_{\rm bol})$
of $1.55\,M_{\odot}$ allows us to determine
$g\alpha$ from 
\begin{equation}
                g\alpha = {L_{\rm bol}\over  M_{\rm IPE\mathhyphen core}\dot E_{\rm ^{56}Ni}(t_{\rm bol}) } \,\, 
\label{eq-IPE-core-mass-b}
\end{equation}
which we obtain using equation~(\ref{eq-IPE-core-mass}) of \S~\ref{section-mass}.
The $g\alpha$ values we obtain are 
$1.33$ ($\rho_{\rm central,WD}=2.9\times10^{9}\,{\rm g\,cm^{-3}}$, $h=1$),
$1.07$ ($\rho_{\rm central,WD}=2.9\times10^{9}\,{\rm g\,cm^{-3}}$, $h=0$),
$1.21$ ($\rho_{\rm central,WD}=10^{10}\,{\rm g\,cm^{-3}}$, $h=1$),
and
$1.00$ ($\rho_{\rm central,WD}=10^{10}\,{\rm g\,cm^{-3}}$, $h=0$).
If we also demand that $g=0.7$ (which as we argued in \S~\ref{section-mass} seems 
a good representative fiducial value), then
the following $\alpha$ values are needed
$1.91$ ($\rho_{\rm central,WD}=2.9\times10^{9}\,{\rm g\,cm^{-3}}$, $h=1$),
$1.53$ ($\rho_{\rm central,WD}=2.9\times10^{9}\,{\rm g\,cm^{-3}}$, $h=0$),
$1.72$ ($\rho_{\rm central,WD}=10^{10}\,{\rm g\,cm^{-3}}$, $h=1$),
and
$1.44$ ($\rho_{\rm central,WD}=10^{10}\,{\rm g\,cm^{-3}}$, $h=0$).
Since $\alpha$ is really not a well constrained parameter, these $\alpha$ values
are not implausible.
Only one is significantly outside of the semi-empirical $\alpha$
range of $\sim 0.6$--$1.7$ we found in \S~\ref{section-comparison-sample}.
Recall that range is not definitive.
Three of the $\alpha$ values are outside of the $\alpha$ range of $0.62$--$1.46$ 
suggested by the SN~Ia models of \citet{hoeflich1996}. 
Recall that range was only suggestive. 
One must remember that SN~2003fg is a very peculiar and a unique SN~Ia, and so its  
$\alpha$ value is not well constrained by the possible $\alpha$ values of 
existing SN~Ia models and other known SNe~Ia. 

     One last fine point to discuss is related to the gamma-ray-escape effect
discussed in \S~\ref{section-mass}.
It follows from the discussion in \S~\ref{section-mass}, that the gamma-ray-escape effect should give
a tendency to lower $\alpha$ as 
$f_{\rm ^{56}Ni}$ 
increases if that
increase brings $^{56}$Ni matter closer to the surface as, in fact, it does in the SSC~model.
From Table~1, we see that the parameter sets with $\alpha=1.2$ have larger 
$f_{\rm ^{56}Ni}$ 
values than the counterpart sets with $\alpha=1.7$ in all cases. 
Thus, coincidentally (since we are not invoking gamma-rays at all in the SSC~model)
$\alpha$ varies with 
$f_{\rm ^{56}Ni}$
in a way that is qualitatively consistent with
tendency of the gamma-ray-escape effect.

\section{CONCLUSIONS AND DISCUSSION\label{section-conclusions}}

        Our analysis of the data for SN~2003fg supports the conclusion of H2006
that this SN~Ia is super-Chandrasekhar (\S~\ref{section-comparison-sn2003fg}).
We find it very improbable, although not absolutely ruled out, that SN~2003fg
has only about a Chandrasekhar mass (the physical Chandrasekhar mass being $1.38\,M_{\odot}$
Appendix~\ref{ap-chan-mass}) and probable that its mass is $\sim 2\,M_{\odot}$
as H2006 also concluded.
Our conclusion relies on the adequacy of YL's binding energy formula for
rotating super-Chandrasekhar-mass WDs (Appendix~\ref{ap-chan-mass}), the analysis of
\citet{langer2000} that limits WD mass in possible accretion histories to $\lesssim  2\,M_{\odot}$,   
and our SSC~model of SN~Ia ejecta (\S~\ref{section-SSC}).
It is likely that improvements to YL's binding energy formula will not change our SSC~model results 
qualitatively, unless
rotating WDs with masses significantly greater than the Chandrasekhar mass are ruled out by improvements
in the input physics of YL's theory of differential rotation in mass-accreting WDs.

        An ultimate verification that SN~2003fg is super-Chandrasekhar would require that only realistic 
explosion models of super-Chandrasekhar-mass WDs allow a radiative transfer history 
(calculated with detailed radiative transfer) consistent with observations.
Since we do not yet have completely adequate explosion models for any SN~Ia, it will probably be some
time before this ultimate verification is possible. 

       What is the significance of SN~2003fg?
Given that as of 2006 September~28 there are 1347 SNe~Ia or possible SNe~Ia that have been observed
according to the \citet{cbat2006} (with, of course,
a wide range of observational coverage from almost nothing to extensive) and 
SN~2003fg is the first SN~Ia discovered for which there is significant evidence of a 
significant super-Chandrasekhar mass, we conclude that SN~2003fg-like SNe~Ia must be rather rare.
Thus, for cosmological evolution and for cosmological distance determinations,  
SN~2003fg-like SNe~Ia are probably of small direct importance.

     But for the study of SNe~Ia,  SN~2003fg may be quite important. 
As YL point out, it is somewhat surprising that super-Chandrasekhar SNe~Ia
have not been much considered given that WDs rotate, and thus can be super-Chandrasekhar
especially if their rotation is accelerated by accretion.   
Of course, the lack of observational need to consider super-Chandrasekhar SNe~Ia
has been a main factor in the relative neglect of such events.  
Also the relative homogeneity of SNe~Ia pointed to a relative of homogeneity progenitors which
pointed to maybe only near-Chandrasekhar mass SNe~Ia.
However, it could be (as YL and \citet{tornambe2005} have discussed at greater length) that     
there is a distribution of SN~Ia progenitor masses from slightly below the 
Chandrasekhar mass to significantly above it and that 
the distribution of masses may explain part of the dispersion of SN~Ia behavior.
This hypothetical distribution is probably fast declining with mass coordinate in the vicinity of
the probable mass of SN~2003fg given that SN~2003fg-like SNe~Ia seem so rare.
In fact, it is possible that having a small super-Chandrasekhar mass is an important
ingredient in normal SNe~Ia explosions and radiative transfer histories. 
This is a speculation, but one that is hard to ignore in light of SN~2003fg.

\acknowledgments

Support for this work has been provided by NASA grant NAG5-3505,
NSF grant AST-0506028,
and
the Homer L. Dodge Department of Physics \&~Astronomy of the University of Oklahoma.
We thank S.-C.~Yoon for answers to questions about his work and Amedeo Tornambe for
stimulating suggestions. 

\appendix

\section{THE CHANDRASEKHAR MASS\label{ap-chan-mass}}

     The expression Chandrasekhar mass is used to mean
the upper limit on the mass of a non-rotating WD. 
Above this upper limit, collapse occurs to a neutron star caused by the inability of
the pressure of the matter's equation of state (EOS) to sustain the WD against its self-gravity.
But what exactly the Chandrasekhar mass is depends on the degree of physical realism specified.

    Chandrasekhar in his book on stellar structure \citep[p.~412ff]{chandrasekhar1957} derives
the upper limit based on Newtonian gravity and the 
zero-temperature, perfect Fermi electron gas in the extreme (special) relativistic limit
where the EOS is
\begin{equation}   
         P=K\rho^{4/3}   \,\, , 
\end{equation}  
where $P$ is pressure, $K$ is a constant depending on fundamental constants and the composition 
of the WD, and $\rho$ is (mass) density.   
It is the electron gas pressure that supports the WD against self-gravity. 
The WD with this EOS is an index-$3$ polytrope with a fixed density profile:  polytropes 
being hydrostatic spheres with EOSs of the form 
\begin{equation}
         P=K\rho^{(n+1)/n}   \,\, ,
\end{equation}
where $K$ is a constant and $n$ is the polytropic index (e.g., \citealt{chandrasekhar1957}, p.~84ff; 
\citealt{clayton1983}, p.~155ff).
From Chandrasekhar's derivation (also given by, e.g., \citet[p.~61--64]{shapiro1983})
with the CODATA 2002 fundamental constant values \citep[e.g.,][]{nist2006a} and the
modern solar mass value $M_{\odot}=1.9891\times10^{33}\,{\rm g}$ \citep[e.g.,][p.~12]{cox2000}, 
one obtains for the mass limit what can be called the formal Chandrasekhar mass,
\begin{equation}   
         M_{\rm ch,formal}={5.8233\over \mu_{e}^{2}} \; M_{\odot}=1.4558 \left({2\over\mu_{e}}\right)^{2} \; M_{\odot} \,\, , 
\label{eq-chan-mass}
\end{equation}  
where the limitation to five significant digits for the coefficients is set by the number of significant digits given for the
modern value of the gravitational constant ($G=6.6742(10)\times10^{-8}$ in CGS units) and
for the modern value of the solar mass. 
(We used double precision arithmetic to calculate the coefficients for equation~(\ref{eq-chan-mass})
and for all the calculations in this appendix.)
The Chandrasekhar mass is independent of density.
Thus, we can imagine scaling up the density throughout the WD by a constant factor and make 
the electron gas more and more relativistic throughout the WD, and thus make the extreme relativistic
condition of derivation more and more exact.
Thus, Chandrasekhar mass is asymptotically exact for zero-temperature, perfect Fermi electron gas
as the density is scaled up toward infinity.
The radius of the WD goes asymptotically to zero as $\rho_{\rm central,WD}\to\infty$
\citep[e.g.,][p.~63]{shapiro1983}. 
Thus, if mass accretion increases a WD mass toward the Chandrasekhar mass, the WD in the original derivation picture is
compacting toward a point mass.
At some point in the accretion, one would expect the input physics to break down and some process would
prevent compaction to a point mass. 

    The $\mu_{e}$ quantity in equation~(\ref{eq-chan-mass}) is the 
mean molecular mass per electron and is given by  
\begin{equation}   
  {1\over\mu_{e}}= \sum_{i}{X_{i}Z_{i}\over A_{i}}   \,\, ,
\end{equation}  
where the sum is over all elements $i$ in the composition, $X_{i}$ is the mass fraction of element $i$,
$Z_{i}$ is the atomic number of element $i$ (we assume that the atoms are all completely ionized and the electrons
all behave as free particles), and $A_{i}$ is the atomic mass of element $i$ in atomic mass units (AMUs)
\citep[e.g.,][p.~84]{clayton1983}.
The effect of composition on the Chandrasekhar mass enters only through $\mu_{e}$.  
For compositions not containing hydrogen or heavy elements $Z_{i}/A_{i}\sim 1/2$, and so the fiducial
value of $\mu_{e}$ is $2$.
For a CO WD with equal parts carbon and oxygen and standard atomic masses (which are some kind of
terrestrial weighted averages of isotopic atomic masses) \citep[e.g.,][]{nist2005},  
$\mu_{e}=2.00085$ and $M_{\rm ch}=1.4546\,M_{\odot}$.

      In evaluating the formal Chandrasekhar mass, Chandrasekhar used older fundamental constant values and
an older solar mass value, of course.
He also used the proton mass where a modern person would probably use the AMU in his formulae, and this
required that his atomic masses for evaluating $\mu_{e}$ are in units of the proton mass rather than in units
of the AMU \citep[p.~415--416,432--433]{chandrasekhar1957}.
For the first coefficient for equation~(\ref{eq-chan-mass}), Chandrasekhar obtained $5.75$ 
\citep[p.~423]{chandrasekhar1957} from which the second coefficient value $1.44$ follows. 
The coefficient values $5.75$ and $1.44$ appear in various places in the literature \citep[e.g.,][]{ostriker1968}.
Using the fundamental constant values and solar mass value ostensibly used by \citet[p.~487]{chandrasekhar1957}
and Chandrasekhar's definition of  $\mu_{e}$, we obtain for the coefficients $5.7513$ and $1.4378$ where we quote more digits
than are physically significant to allow consistency checks:  Chandrasekhar correctly rounded off to 3 digits
given the precision of the values he used.
If one does the calculations for the coefficients using the EOS parameter values calculated by
\citet[p.~416]{chandrasekhar1957}
and his gravitational constant ($G=6.67\times10^{-8}$ in CGS units:  \citealt[p.~487]{chandrasekhar1957}) 
rather than using 
the older fundamental constant values (except for the gravitational constant) directly (which implies
using Chandrasekhar's definition of $\mu_{e}$) and using the modern solar mass value,
one obtains for the coefficients $5.7444$ and $1.4361$ where we quote more digits   
than are physically significant to allow consistency checks.
If we used the modern fundamental constant values and solar mass value and Chandrasekhar's definition of  $\mu_{e}$, 
the coefficients in equation~(\ref{eq-chan-mass}) would be $5.7395$ and $1.4349$ to the correct number of 
significant digits.
We report the numbers in this paragraph to illustrate some of the sources of small inconsistencies in
formal Chandrasekhar mass values that may appear in the literature. 

      The formal Chandrasekhar mass is an idealization in several respects. 
If one makes a correction for general relativity, then there is a collapse caused by gravity that occurs
at a finite density which when considered apart from other effects is $2.646\times10^{10}(\mu_{e}/2)^{2}\,{\rm g\,cm^{-3}}$
\citep[e.g.,][p.160]{shapiro1983} and at mass slightly different from that of  equation~(\ref{eq-chan-mass}).
Since the collapse occurs at a finite density, the electron gas is not in the extreme relativistic limit
throughout the WD and gives rise to another correction for the collapse mass. 
The two effects reduce the mass of collapse from $M_{\rm ch,formal}$ to 
$1.418 \left({2/\mu_{e}}\right)^{2}\,M_{\odot}$ \citep[e.g.,][p.~84]{woosley1994}. 
There is also a Coulomb correction to account for the concentration of positive charge into nuclei
rather spread uniformly in the electron gas \citep[e.g.,][p.~29ff]{shapiro1983}. 
Adding this effect gives a corrected Chandrasekhar mass $M_{\rm ch}$:
\begin{eqnarray}   
  M_{\rm ch}&\approx& 1.418 \left({2\over\mu_{e}}\right)^{2}
               \left[1-{3\over5}\left({12\over\pi}\right)^{1/3}\alpha_{\rm fs}
                                           \langle Z\rangle^{2/3}\right] \; M_{\odot}    \cr
\noalign{\smallskip} 
      &\approx& 1.418 \left({2\over\mu_{e}}\right)^{2}\left[1-0.02259927661
                             \times\left({\langle Z\rangle\over 6}\right)^{2/3}\right] \; M_{\odot} 
\end{eqnarray} 
(\citealt{shapiro1983}, p.~32;  \citealt{baron1990}; \citealt{woosley1994}, p.~84),
where $\alpha_{\rm fs}=1/137.03599911(46)$ is the fine structure constant \citep[e.g.,][]{nist2006a}
and
where we have approximately generalized the expression from the references
by replacing atomic number $Z$ by mean atomic number $\langle Z\rangle$. 
For a CO WD with equal parts carbon and oxygen, $\mu_{e}=2.00085$ again, and $\langle Z\rangle=7$, one obtains
$M_{\rm ch}=1.38\,M_{\odot}$.

     Further corrections to the corrected Chandrasekhar mass are needed for finite temperature
(\citealt{baron1990}; \citealt{woosley1994}, p.~85--86).
These corrections increase the corrected Chandrasekhar mass, but for CO WDs they do so by only 
of order $0.1\,$\%.  

     In fact, it is believed that non-rotating CO WDs can never reach the exact corrected Chandrasekhar mass 
(whatever that exactly is) for a density-insensitive EOS. 
It is believed that WDs can only grow close to the corrected Chandrasekhar mass by accretion.
During this process when central density $\rho_{\rm central,WD}$ reaches of order $10^{10}\,{\rm g\,cm^{-3}}$, electron capture
on nuclei to make neutrons is likely to induce collapse to a neutron star by rapidly diminishing the electron gas pressure
(\citealt{nomoto1991};  YL).
This collapse event depends on the central temperature being low enough, but that seems likely in likely accretion histories.
If the central temperature is relatively high (of order $2.5\times10^{8}\,$K), then unstable carbon burning is likely
to lead to thermonuclear explosion when the central density is of order $3\times10^{9}\,{\rm g\,cm^{-3}}$
\citep[e.g.,][p.~108]{woosley1994}.
The central density and temperature for collapse or explosion actually do depend on the accretion history and are also
are subject to revision with improved input physics.
However, at present the non-rotating WD mass at the time of explosive carbon ignition is found to be about $1.38\,M_{\odot}$ 
(e.g., \citealt{woosley1994}, p.~108;  \citealt{nomoto1984}):  i.e., close to the zero-temperature
corrected Chandrasekhar mass $M_{\rm ch}$.  
In the main text, we will just refer to $1.38\,M_{\odot}$ as 
the Chandrasekhar mass or physical Chandrasekhar mass for brevity.

     As a necessary simplification for their rotating WD modeling, YL adopted the  
zero-temperature, perfect Fermi electron gas EOS derived by
Chandrasekhar in 1935 \citep{chandrasekhar1935}.
This EOS is sometimes called the Chandrasekhar EOS \citep[e.g.,][]{ostriker1968}.
The Chandrasekhar EOS is more advanced than that used to derive the formal Chandrasekhar mass 
in that it allows for the full range of special relativistic effects from none to extreme.
YL actually use the parameter values for the Chandrasekhar EOS that are given 
Chandrasekhar's book \citep[p.~416]{chandrasekhar1957} and adopt $\mu_{e}=2$ for all calculations.
By implication, YL should be using Chandrasekhar's definition of $\mu_{e}$.
With this definition, $\mu_{e}$ should be 1.98640 for equal amounts of carbon and oxygen
(calculated using standard atomic masses times the AMU divided by the proton mass), not 2.
YL's value for the Chandrasekhar mass for their non-rotating model with infinite central
density is $1.436\,M_{\odot}$ which is just the Chandrasekhar mass value one would
calculate using Chandrasekhar's EOS parameter values, Chandrasekhar's value of the gravitational
constant, the modern solar mass value, Chandrasekhar's definition of $\mu_{e}$, 
and $\mu_{e}=2$.  
YL's Chandrasekhar mass value was, in fact, determined as just described \citep{yoon2006}.

      Because YL have omitted the general relativistic, Coulomb, and temperature corrections discussed
above to the EOS, their masses will likely be in error by a few percent for the given input parameters. 
The Coulomb correction is probably the most important omission. 
It tends to reduce mass for a given central density.
Thus, YL's masses are probably overestimates for their specified central densities. 
Actually, YL's treatment of the EOS (using older parameter values for the EOS and $\mu_{e}=2$ as described
just above) may partially compensate for omitting the Coulomb correction.
For example, the Coulomb correction reduces the Chandrasekhar mass by about $2.5\,$\% and
and YL's treatment reduces the formal Chandrasekhar mass for
equal amounts of carbon and oxygen from $1.4546\,M_{\odot}$ (see above) to $1.436\,M_{\odot}$ or by about $1.3\,$\%.

\section{THE YOON \&~LANGER BINDING ENERGY FORMULAE FOR ROTATING SUPER-CHANDRASEKHAR 
                      CO WHITE DWARFS AND SOME RELATED RESULTS\label{ap-YL-formula}}

     The YL formula for the binding energy of rotating super-Chandrasekhar CO WDs is
\begin{eqnarray}
   E_{\rm bind}(\rho_{\rm central,WD},M)
     &=&E_{\rm bind,NR}(\rho_{\rm central,WD})      \cr
     &&\qquad   +E_{\rm bind,coef}(\rho_{\rm central,WD})\left[M-M_{\rm NR}(\rho_{\rm central,WD})\right]^{1.03} 
\label{eq-energy-bind}
\end{eqnarray}
(YL's eq.~(32)), 
where the energy units are foes
(i.e., units of $10^{51}$ ergs),
$\rho_{\rm central,WD}$ is the WD central density in grams per centimeter cubed,
$M$ is the WD mass in solar mass units,
\begin{eqnarray}
  E_{\rm bind,NR}(\rho_{\rm central,WD})&=&0.1\times\Biggl\{
                  -32.759747
                  +6.7179802\times\log(\rho_{\rm central,WD})   \cr
                 && \qquad\qquad -0.28717609\times[\log(\rho_{\rm central,WD})]^{2} \Biggr\}
\label{eq-energy-bind-nr}
\end{eqnarray}
is the binding energy in foes for a non-rotating WD with central density $\rho_{\rm central,WD}$ (YL's eq.~(34)),
\begin{eqnarray}
  E_{\rm bind,coef}(\rho_{\rm central,WD})&=&0.1\times\Biggl\{
                    -370.73052
                    +132.97204\times\log(\rho_{\rm central,WD})   \cr 
                  &&\qquad\qquad   -16.117031\times[\log(\rho_{\rm central,WD})]^{2}   \cr
                  &&\qquad\qquad\qquad  +0.66986678\times[\log(\rho_{\rm central,WD})]^{3} \Biggr\}
\end{eqnarray}
in foes (YL's eq.~(33)), and
\begin{eqnarray}
M_{\rm NR}(\rho_{\rm central,WD})&=&1.436\times
    \Biggl\lfloor 1-\exp\Biggl\{-0.01316\times[\log(\rho_{\rm central,WD})]^{2.706} \cr
    &&\qquad\qquad\qquad\qquad\qquad     +0.2493\times\log(\rho_{\rm central,WD})
                      \Biggr\}
    \Biggr\rfloor 
\label{eq-mass-nr}
\end{eqnarray}
is the mass in solar mass units of a non-rotating CO WD with central density $\rho_{\rm central,WD}$
(YL's eq.~(22)). 
The binding energy formula is a fit to the numerical results of YL and is verified for central densities
in the range $10^{8}$--$10^{10}\,{\rm g\,cm^{-3}}$ and
mass in the range  $\sim 1.16$--$2.1\,M_{\odot}$.
Recall from \S~\ref{section-mass} that the CO WD central density range for explosion as SNe~Ia is 
$\sim 2.5\times10^{9}$--$10^{10}\,{\rm g\,cm^{-3}}$ (\citealt{woosley2006}; \citealt{nomoto1991};  YL). 
We have extrapolated the use of the binding energy formula up to $3\,M_{\odot}$.

    The formula for $M_{\rm NR}(\rho_{\rm central,WD})$ is verified for $\rho_{\rm central,WD}>10^{7}\,{\rm g\,cm^{-3}}$:
for $\rho_{\rm central,WD}=2.5\times10^{9}\,{\rm g\,cm^{-3}}$, it gives $1.384\,M_{\odot}$ and 
for $\rho_{\rm central,WD}=10^{10}\,{\rm g\,cm^{-3}}$, $1.414\,M_{\odot}$.
Note that as $\rho_{\rm central,WD}$ goes to infinity, $M_{\rm NR}(\rho_{\rm central,WD})$ goes to $1.436\,M_{\odot}$.
The value $1.436\,M_{\odot}$ is the formal Chandrasekhar mass for YL's choice of equation of state and its parameters
(see Appendix~\ref{ap-chan-mass}).
For the purposes of the investigations of this paper, the differences between $1.436\,M_{\odot}$ and
the physical Chandrasekhar mass of $1.38\,M_{\odot}$ (see Appendix~\ref{ap-chan-mass}) and between 
YL's equation of state and the exact equation of state are small.

     The binding energy formula is, in fact, almost linear in $M$ for a fixed $\rho_{\rm central,WD}$ 
(see YL's Fig.~9).
Doing least-squares fits of the binding energy formula to a line for
the mass range $\sim 1.4$--$3\,M_{\odot}$ for the limiting central densities for SNe~Ia give
\begin{equation}
E_{\rm bind}(\rho_{\rm central,WD}=2.5\times10^{9}\,{\rm g\,cm^{-3}},M)=(-1.133\pm0.008)+(1.169\pm0.003)\times M
\end{equation}
and
\begin{equation}
E_{\rm bind}(\rho_{\rm central,WD}=10^{10}\,{\rm g\,cm^{-3}},M)=(-1.923\pm0.011)+(1.746\pm0.005)\times M \,\, , 
\end{equation}
where again the energy is in foes and the mass in solar mass units. 
For exploring the behavior of the
formula for kinetic energy of a SNe~Ia (eq.~(\ref{eq-energy-b}) in \S~\ref{section-SSC}), it is, in fact, convenient
to use a linear fit to the binding energy. 
With such a fit, the kinetic energy formula is 
\begin{equation}
         E \approx E_{\rm IPE}M[f_{\rm IPE}+Rh(1-f_{\rm IPE})] - a - bM \,\, ,
\label{eq-energy-c}
\end{equation}
where from the above we know that intercept $a$ and slope $b$ vary with allowed central densities for SNe~Ia over
$-1.133$ to $-1.923$ and $1.169$ to $1.746$, respectively. 

     One use of the linear approximation to the binding energy formula is to verify 
that $\sqrt{2E/M}$ is roughly constant for fixed fractions of explosion-synthesized elements.
We made use of this constancy in \S~\ref{section-mass} in our simple scaling argument for SN~Ia mass.
Substituting in $E_{\rm IPE}=1.61\,{\rm foe}/M_{\odot}$ and $R=0.768$ (see \S~\ref{section-mass})
into equation~(\ref{eq-energy-c})
and dividing that equation by $M$ gives
\begin{equation}
         {E\over M}\approx 1.61\times[f_{\rm IPE}+0.768\times h(1-f_{\rm IPE})]-b+{|a|\over M} \,\, . 
\label{eq-e-over-m-a}
\end{equation}
To have any kind of explosion at all $E/M>0$, and so the constant term of
equation~(\ref{eq-e-over-m-a}) must be $\gtrsim -|a_{\rm min}|/M_{\rm ch}\approx -1.9/1.4\approx -1.4$  
(where $a_{\rm min}=-1.923$ is the minimum value of $a$ for the $\rho_{\rm central,WD}$ allowed range). 
If we choose $f_{\rm IPE}=0.648$, $h=0.798$, 
and $\rho_{\rm central,WD}=2.6\times10^{9}\,{\rm g\,cm^{-3}}$ (which are the values of model~W7
\citep{nomoto1984,thielemann1986}) as representative of SN~Ia explosions, then 
$a=-1.15$ and $b=1.18$
(from a least-squares fit to the binding energy formula over the mass range $\sim 1.4$--$3\,M_{\odot}$ with
$\rho_{\rm central,WD}=2.6\times10^{9}\,{\rm g\,cm^{-3}}$) 
and the first term of equation~(\ref{eq-e-over-m-a}) is $1.39$.
(Note $f_{\rm IPE}=0.648$ and $h=0.798$, imply  
$f_{\rm CO}=0.0711$ and $f_{\rm IME}=0.281$ using equation~(\ref{eq-fco-fime}) in \S~\ref{section-SSC}.)
Substituting the calculated values for $a$, $b$, and the first term of equation~(\ref{eq-e-over-m-a})
into equation~(\ref{eq-e-over-m-a}) gives
\begin{equation}
         {E\over M}\approx 0.21+{1.15\over M} \,\, .
\label{eq-e-over-m-b}
\end{equation}
Thus, with the representative values, an explosion can be expected for any $M$ since the constant term 
of equation~(\ref{eq-e-over-m-b}) is positive, but the value of $E/M$ can only be very crudely 
approximated as a constant with respect to $M$.
However, $\sqrt{2E/M}$ will be closer to constancy and we can accept that it is so crudely.
(Making use of equation~(\ref{eq-e-over-m-b}) and converting units, we find for 
$M$ going from $1.38\,M_{\odot}$ to $3\,M_{\odot}$ that $\sqrt{2E/M}$
goes from $10260\,{\rm km\,s^{-1}}$ to $7750\,{\rm km\,s^{-1}}$.)
This constancy verifies one of our assumptions for the simple scaling argument
provided that the SSC~model is valid up to $\sim 3\,M_{\odot}$
which requires that the usage of YL's binding energy formula can be extrapolated to $3\,M_{\odot}$. 

\section{SOLVING THE SSC MODEL FOR MASS TREATED AS A DEPENDENT PARAMETER\label{ap-SSC-mass}}

      To calculate Figures~\ref{f1} and~\ref{f2}, we used the SSC model mass $M$ as an independent parameter.
(See \S\S~\ref{section-mass} and~\ref{section-SSC} for a description of the parameters and the SSC model.)
One can make $M$ a dependent parameter and replace it among the independent parameters with $v_{\rm IPE\mathhyphen core}$.
The solution for $M$ can then be done via the Newton-Raphson method \citep[e.g.,][p.~355ff]{press1992}
with IPE-core mass $M_{\rm IPE\mathhyphen core}$, IPE-core velocity $v_{\rm IPE\mathhyphen core}$ (now independent), central
density $\rho_{\rm central,WD}$, and $h$ as the 4 independent parameter values.
We assume the independent parameter values are given.
The dependent parameters besides $M$ are 
the $e$-folding velocity $v_{e}$, kinetic energy $E$, and the mass fractions $f_{\rm CO}$, $f_{\rm IME}$, and $f_{\rm IPE}$.

     The Newton-Raphson method starts with an initial input mass $M_{1}$ (for which
we suggest an initial value below) and iterates to improve the input values toward the true value for $M$.
The steps of the iteration are as follows. 
An input $M_{i}$ for the $i$th iteration along with the given $M_{\rm IPE\mathhyphen core}$, $\rho_{\rm central,WD}$ and $h$ values
are used to evaluate the following version 
of the kinetic energy formula (see eq.~(\ref{eq-energy-b})
in \S~\ref{section-SSC}): 
\begin{equation}
         E=E_{\rm IPE}[M_{\rm IPE\mathhyphen core}+Rh(M-M_{\rm IPE\mathhyphen core})] - E_{\rm bind}(\rho_{\rm central,WD},M)  \,\, ,
\label{eq-energy-d}
\end{equation}
where we have used the fact that $M_{\rm IPE\mathhyphen core}=Mf_{\rm IPE}$ (see eq.~(\ref{eq-fipe-fime}) in \S~\ref{section-mass})
to eliminate $f_{\rm IPE}$ from equation~(\ref{eq-energy-b}).
Using the calculated $i$th iteration kinetic energy $E_{i}$, we obtain the $i$th iteration $e$-folding velocity
\begin{equation}
         v_{e,i}=\sqrt{{B\over6}{{E_{i}\over M_{i}}}} \,\, , 
\label{eq-efolding-b}
\end{equation}
where we have used equation~(\ref{eq-efolding}) (\S~\ref{section-SSC}), but with
assumption that $E_{i}$ is in foes and $M_{i}$ is in solar mass units, and so need conversion factor
$B=10^{51}\,{\rm ergs\,foe^{-1}}/M_{\odot}$ to give $v_{e,i}$ in CGS units.
The $i$th iteration IPE-core reduced velocity 
$x_{{\rm IPE\mathhyphen core},i}=v_{\rm IPE\mathhyphen core}/v_{e,i}$ (see eq.~(\ref{eq-reduced-velocity})
in \S~\ref{section-SSC}) (assuming the given $v_{\rm IPE\mathhyphen core}$ value is in CGS units).
One then uses equation~(\ref{eq-interior-mass-fraction}) (\S~\ref{section-SSC}), to find the $i$th iteration
$f_{\rm IPE,i}$ and then
the $i$th iteration output mass $M_{{\rm out},i}$ of the iteration follows from
\begin{equation}
         M_{{\rm out},i}={M_{\rm IPE\mathhyphen core}\over f_{\rm IPE,i}} \,\, , 
\end{equation}
where we have made use of equation~(\ref{eq-fipe-fime}) in (\S~\ref{section-mass})
For the Newton-Raphson method iteration, the quantity whose zero we want to find as a function of $M$ is $M_{\rm out}-M$.
The $i$th iteration Newton-Raphson correction $\Delta M_{i}$ to approximately zero $M_{\rm out}-M$ is given by
\begin{equation}
        \Delta M_{i}= -\left({M_{\rm out}-M\over dM_{\rm out}/dM-1}\right)\Bigg|_{M_{i}}  \,\, .
\label{eq-newton-raphson-correction}
\end{equation}
It is straightforward to show that the $i$th iteration value of the derivative $dM_{\rm out}/dM$ is given by 
\begin{eqnarray}
  {dM_{\rm out}\over dM}\Bigg|_{M_{i}}
&=&-\left({M_{\rm IPE\mathhyphen core}\over f^{2}}{df\over dx}{dx\over dv_{e}}{dv_{e}\over dw}{dw\over dM}\right)\Bigg|_{M_{i}} \cr
&=&\left({M_{\rm IPE\mathhyphen core}\over f_{\rm IPE,i}^{2}}\right)
                            \left[{x_{{\rm IPE\mathhyphen core},i}^{3}\exp(-x_{{\rm IPE\mathhyphen core},i})\over24}\right]
                            \left({B\over v_{e,i}^{2}}\right)\left[{d(E/M)\over dM}\right]\Bigg|_{M_{i}} \,\, ,
\label{eq-mass-out}
\end{eqnarray}
where $w=(B/6)(E/M)$,
\begin{equation}
  {d(E/M)\over dM} = {E_{\rm IPE}Rh-{dE_{\rm bind}/dM} \over M} - {E\over M^{2}}  \,\, ,
\end{equation}
and
\begin{equation}
  {dE_{\rm bind}\over dM}= 
     1.03\times E_{\rm bind,coef}(\rho_{\rm central,WD})\left[M-M_{\rm NR}(\rho_{\rm central,WD})\right]^{0.03}  \,\, .
\end{equation}
The last equation was obtained by differentiation from equation~(\ref{eq-energy-bind}). 

      The Newton-Raphson method in this case converges quite quickly in practice.
The guarantee of convergence is that $M_{\rm out}-M$ is strictly decreasing with respect to $M$ with
no stationary points for the allowed $\rho_{\rm central,WD})$ range 
for SN~Ia explosion of
$\sim 2.5\times10^{9}$--$10^{10}\,{\rm g\,cm^{-3}}$ (\citealt{woosley2006}; \citealt{nomoto1991};  YL).
This means that the Newton-Raphson correction $\Delta M_{i}$ always has the correct sign
and if the corrections are prevented from causing any overshooting of the diminishing allowed range for
the true mass $M$, then convergence must follow.
All that is required to show the strictly decreasing nature of $M_{\rm out}-M$ is to
show that ${E/M}$ is monotonically
decreasing with respect to $M$ with a stationary point only at $M=\infty$.
This is clear from equation~(\ref{eq-mass-out}) since all the factors on the right-hand side of the second expression
are clearly always positive or zero except for $d(E/M)/dM$.
(Note that $M_{\rm out}-M$ has no stationary points even though $d(E/M)/dM$ and $dM_{\rm out}/dM$ do because 
the $-M$ term derivative is $-1$.)

     Because of the unusual exponent in the binding energy formula (see eq.~(\ref{eq-energy-bind}) in
Appendix~\ref{ap-YL-formula}), it is awkward and not enlightening to work with the exact formula
for kinetic energy equation~(\ref{eq-energy-d}).
Instead we will show that ${E/M}$ is monotonically decreasing with respect to $M$ using the kinetic
energy formula with the linear binding energy formula approximation:  see
equation~(\ref{eq-energy-c}) in Appendix~\ref{ap-YL-formula}.
First, we rewrite equation~(\ref{eq-energy-c})
in terms of the proper variables for the Newton-Raphson method: 
\begin{eqnarray}
         E &\approx& E_{\rm IPE}[M_{\rm IPE\mathhyphen core}+Rh(M-M_{\rm IPE\mathhyphen core})] - a - bM  \cr
                 &=& E_{\rm IPE}M_{\rm IPE\mathhyphen core}(1-Rh)-a +(E_{\rm IPE}Rh-b)M \,\, , 
\label{eq-energy-e}
\end{eqnarray}
where we note that the intercept term $E_{\rm IPE}M_{\rm IPE\mathhyphen core}(1-Rh)-a>0$ always since $1-Rh>0$ always and $a<0$
for all linear fits to the binding energy formula for its allowed $\rho_{\rm central,WD})$ range
(see Appendix~\ref{ap-YL-formula}). 
The sign of the slope $E_{\rm IPE}Rh-b$ depends on the relative sizes of $h$ and $b$.
Using the model~W7 \citep{nomoto1984,thielemann1986} values $h=0.798$ and $b=1.18$, we find that
$E_{\rm IPE}Rh-b=-0.19$:  thus, what we can take as the representative slope of
equation~(\ref{eq-energy-e}) is negative.
Using equation~(\ref{eq-energy-e}), we obtain
\begin{equation}
        {E\over M} \approx {E_{\rm IPE}M_{\rm IPE\mathhyphen core}(1-Rh)-a \over M} + E_{\rm IPE}Rh-b \,\, ,
\label{eq-e-over-m-c}
\end{equation}
and then differentiating $E/M$ we get
\begin{equation}
        {d(E/M)\over dM} \approx - \left[{E_{\rm IPE}M_{\rm IPE\mathhyphen core}(1-Rh)-a \over M^{2}}\right]  \,\, .
\end{equation}
Since $E_{\rm IPE}M_{\rm IPE\mathhyphen core}(1-Rh)-a>0$ always, it
is clear within our approximation for $E$ that ${E/M}$ is monotonically decreasing with the only stationary point
at $M=\infty$.
Now we have shown this behavior only for the approximate equation~(\ref{eq-e-over-m-c}).
But since this equation is, in fact, a very good approximation to equation~(\ref{eq-energy-d}) 
divided by $M$ and the behavior is robust for all allowed 
values of $a$ 
(see Appendix~\ref{ap-YL-formula}),
we conclude that ${E/M}$ is to at least good accuracy monotonically decreasing with $M$ in general with a stationary point only
at $M=\infty$. 
It then follows from equation~(\ref{eq-mass-out}) that $M_{\rm out}-M$ is to at least good
accuracy strictly decreasing with $M$.
This is what we needed to show to prove that
the Newton-Raphson method would always converge for the allowed $\rho_{\rm central,WD})$ range.  

     We can
deduce some approximate limiting behaviors of the $M_{\rm out}$ function by making use of equation~(\ref{eq-e-over-m-c}).
If $E_{\rm IPE}Rh-b\leq 0$ (which we take to be the representative case) and 
$M$ is increased, then at some point (which is at finite $M$ if 
$E_{\rm IPE}Rh-b<0$ and infinite $M$ otherwise) 
$E/M\to0$, $v_{e}\to0$, $x\to\infty$, $f\to1$, and $M_{\rm out}\to M_{\rm IPE\mathhyphen core}$. 
On the other hand, if $E_{\rm IPE}Rh-b>0$ and $M\to\infty$, then
$E/M$, $v_{e}$, $x$, $f$, and $M_{\rm out}$ approach asymptotically to a finite positive value,
a finite positive value, a finite positive value, a finite positive value less than 1, and
a finite positive value greater than $M_{\rm IPE\mathhyphen core}$, respectively. 
If $M\to0$, then $E/M\to\infty$, $v_{e}\to\infty$, $x\to0$, $f\to0$, and $M_{\rm out}\to\infty$.
Actually, we cannot let $M\to0$ in the $M_{\rm out}$ function since YL's binding energy formula
has no solution for $M<M_{\rm NR}$.

        What is the appropriate initial value $M_{1}$ for the Newton-Raphson method?
Well the lower bound on the true value for $M$ is clearly  
\begin{equation}
             M_{\rm lower\;bound}=\max\left[M_{\rm IPE\mathhyphen core},M_{\rm NR}(\rho_{\rm central,WD})\right] \,\, 
\end{equation} 
since a physical $M$ cannot be less than the input core mass $M_{\rm IPE\mathhyphen core}$ nor than the mass of a non-rotating
WD for the input density $M_{\rm NR}(\rho_{\rm central,WD})$. 
In practice, we have found $M_{\rm lower\;bound}$ to work well as an initial value $M_{1}$.

    If we use $M_{\rm lower\;bound}$ as an input mass $M$ and find that $M_{\rm out}(M_{\rm lower\;bound})<M_{\rm lower\;bound}$, 
then there is no solution since Newton-Raphson correction from equation~(\ref{eq-newton-raphson-correction}) will
be negative.
Recall that Newton-Raphson correction always has the correct sign for our case, and so a negative correction to the
lower bound on $M$ implies there is no solution. 
One way of looking at this case is to say there is not enough nuclear burning energy released with any allowed value
of $M$ to match the input $v_{\rm IPE\mathhyphen core}$.  
On Figures~\ref{f1} and~\ref{f2}, the no-solution case with
$M_{\rm lower\;bound}=M_{\rm NR}(\rho_{\rm central,WD})$ and
$M_{\rm out}(M_{\rm lower\;bound})<M_{\rm NR}(\rho_{\rm central,WD})$
would correspond to having the point $(M_{\rm IPE\mathhyphen core},v_{\rm IPE\mathhyphen core})$ in the region to the left
of the $M_{\rm NR}$ mass curves:  if was to the right, there would be a solution.
This no-solution case is actually the case for the point~N on Figure~\ref{f2} for
$h=0$ (which means $f_{\rm IME}=0$ recall) (see \S~\ref{section-comparison-sn2003fg}).

    If $M_{\rm lower\;bound}$ gives $E<0$, then there is also no solution mass $M$.
Since the intercept term of equation~(\ref{eq-energy-e}) is always positive, then $E<0$ for
any input $M$ implies the slope is negative.
This means that the approximate kinetic energy function (and the exact kinetic energy function to high accuracy)
can only grow more negative for $M$ increasing beyond
$M_{\rm lower\;bound}$.
Thus for no allowed $M$ is the kinetic energy positive which is required for an explosion.
There is not enough nuclear burning energy for any allowed $M$ to explode the WD
with the input parameter values in this case.

     The true values for $f_{\rm IPE}$, $v_{e}$, kinetic energy $E$, and mass $M$ are obtained from the Newton-Raphson iteration.
The true values of $f_{\rm CO}$ and $f_{\rm IME}$ are obtained using the $h$ and $f_{\rm IPE}$ values in
equation~(\ref{eq-fco-fime}) (\S~\ref{section-SSC}).

     Note that if we had used either of $f_{\rm CO}$ or $f_{\rm IME}$ as an independent parameter instead
of $h$, we would have to deal with an extra check on whether or not a given converged solution $M$ was physically
allowed.
If we used $f_{\rm CO}$ ($f_{\rm IME}$) as an independent parameter and solved for $M$ and then found 
that $f_{\rm IME}<0$ ($f_{\rm CO}<0$), the found solution $M$ would be ruled out:
i.e., there would be no physically allowed solution for the given input parameter values.  
Using $h$ as an independent parameter eliminates this complication for valid $h$ (i.e., $h$ in the range $0$--$1$) and
a solution mass $M\geq M_{\rm lower\;bound}$, since one will only obtain valid $f_{\rm CO}$ and $f_{\rm IME}$ values
from equation~(\ref{eq-fco-fime}) (\S~\ref{section-SSC}):  i.e., values in the range $0$--$1$.
Thus, the solution mass $M$ can be known to be valid without having to check on the $f_{\rm CO}$ or $f_{\rm IME}$ values. 
The above discussion shows why we prefer using $h$ as an independent parameter to either of $f_{\rm CO}$ or $f_{\rm IME}$.

\section{SOLVING THE SSC MODEL FOR CORE MASS TREATED AS A DEPENDENT PARAMETER\label{ap-SSC-IPE-core-mass}}

      In this appendix, we show how to solve the SSC~model for core mass $M_{\rm IPE\mathhyphen core}$ treated
as a dependent parameter.
The independent parameters are now mass $M$, IPE-core velocity $v_{\rm IPE\mathhyphen core}$, WD central density
$\rho_{\rm central,WD}$, and $h$.
We assume the independent parameter values are given.
The dependent parameter besides $M_{\rm IPE\mathhyphen core}$ are
the $e$-folding velocity $v_{e}$, kinetic energy $E$, and mass fractions $f_{\rm CO}$, $f_{\rm IME}$, and $f_{\rm IPE}$.
The solution is by the Newton-Raphson method \citep[e.g.,][p.~355ff]{press1992} 
and is similar to that of Appendix~\ref{ap-SSC-mass},
but is somewhat simpler since the binding energy formula (eq.~(\ref{eq-energy-bind}) in Appendix~\ref{ap-YL-formula})
does not depend on $M_{\rm IPE\mathhyphen core}$.   

     The Newton-Raphson method iteration starts with an initial input core mass $M_{\rm IPE\mathhyphen core,1}$ (for which
we suggest an initial value below) and iterates to improve the input values toward the true value for $M_{\rm IPE\mathhyphen core}$.
The steps of the iteration are as follows.
An input $M_{{\rm IPE\mathhyphen core},i}$ for the $i$th iteration is used to evaluate the $i$th iteration
kinetic energy and $e$-folding velocity $v_{e,i}$ using,
respectively, equations~(\ref{eq-energy-d}) and~(\ref{eq-efolding-b}).
The $i$th iteration IPE-core reduced velocity $x_{{\rm IPE\mathhyphen core},i}=v_{\rm IPE\mathhyphen core}/v_{e,i}$ 
(see eq.~(\ref{eq-reduced-velocity}) in \S~\ref{section-SSC})
(assuming the given $v_{\rm IPE\mathhyphen core}$ value is in CGS units).
One then uses equation~(\ref{eq-interior-mass-fraction}) (\S~\ref{section-SSC}), to find the $i$th iteration
$f_{{\rm IPE},i}$ and then the $i$th iteration output core mass $M_{{\rm IPE\mathhyphen core,out},i}$ follows from
\begin{equation}
         M_{{\rm IPE\mathhyphen core,out},i}=M f_{{\rm IPE},i} \,\, , 
\end{equation}
where we have used equation~(\ref{eq-fipe-fime}) (\S~\ref{section-mass}).
For the Newton-Raphson method iteration, the quantity whose zero we want to find as a function of 
$M_{\rm IPE\mathhyphen core}$ is $M_{\rm IPE\mathhyphen core,out}-M_{\rm IPE\mathhyphen core}$.
The $i$th iteration Newton-Raphson correction $\Delta M_{{\rm IPE\mathhyphen core},i}$ 
to approximately zero $M_{\rm IPE\mathhyphen core,out}-M_{\rm IPE\mathhyphen core}$ is given by
\begin{equation}
        \Delta M_{{\rm IPE\mathhyphen core},i}
     = -\left({M_{\rm IPE\mathhyphen core,out}-M_{\rm IPE\mathhyphen core}
           \over dM_{\rm IPE\mathhyphen core,out}/dM_{\rm IPE\mathhyphen core}-1}\right)
             \Bigg|_{M_{{\rm IPE\mathhyphen core},i}} \,\, .
\label{eq-newton-raphson-correction-b}
\end{equation}
It is straightforward to show that the $i$th iteration derivative 
$dM_{\rm IPE\mathhyphen core,out}/dM_{\rm IPE\mathhyphen core}$ is given by 
\begin{eqnarray}
        {dM_{\rm IPE\mathhyphen core,out}\over dM_{\rm IPE\mathhyphen core}}\Bigg|_{M_{{\rm IPE\mathhyphen core},i}}
      &=& \left( M{df\over dx}{dx\over dv_{e}}{dv_{e}\over dw}
                  {dw\over dM_{\rm IPE\mathhyphen core}}\right)\Bigg|_{M_{{\rm IPE\mathhyphen core},i}}   \cr
      &=&                 -\left[{x_{{\rm IPE\mathhyphen core},i}^{3}\exp(-x_{{\rm IPE\mathhyphen core},i})\over24}\right] \cr
      && \qquad\qquad     \times   \left({B\over v_{e,i}^{2}}\right)E_{\rm IPE}(1-Rh)     \,\, , 
\end{eqnarray}
where $w=(B/6)(E/M)$.
The derivative $dM_{\rm IPE\mathhyphen core,out}/dM_{\rm IPE\mathhyphen core}\leq0$ always with the equality holding only
for $x_{{\rm IPE\mathhyphen core},i}=0$ and $x_{{\rm IPE\mathhyphen core},i}=\infty$.
(Note $x_{{\rm IPE\mathhyphen core},i}^{3}/v_{e,i}^{2}=x_{{\rm IPE\mathhyphen core},i}^{5}/v_{\rm IPE\mathhyphen core}^{2}$.)
Thus, the Newton-Raphson method is guaranteed to converge to the true $M_{\rm IPE\mathhyphen core}$ value
since $M_{\rm IPE\mathhyphen core,out}-M_{\rm IPE\mathhyphen core}$
is strictly decreasing with $M_{\rm IPE\mathhyphen core}$ provided the Newton-Raphson corrections
are prevented from causing any overshooting of the diminishing allowed range for
the converged solution and
provided the independent parameter values allow a valid $M_{\rm IPE\mathhyphen core}$ solution (i.e., one in the range $0$--$M$). 
(Note that $M_{\rm IPE\mathhyphen core,out}-M_{\rm IPE\mathhyphen core}$ has no stationary points even though 
$M_{\rm IPE\mathhyphen core,out}$ does 
because the $-M_{\rm IPE\mathhyphen core}$ term derivative is $-1$.)

     The true $f_{\rm IPE}$, $e$-folding velocity $v_{e}$, kinetic energy $E$, 
and mass $M_{\rm IPE\mathhyphen core}$ values are obtained from the iteration.
The true values of $f_{\rm CO}$ and $f_{\rm IME}$ are obtained using the $h$ and $f_{\rm IPE}$ values in
equation~(\ref{eq-fco-fime}) (\S~\ref{section-SSC}).

      A good initial input $M_{\rm IPE\mathhyphen core,1}$ for Newton Raphson iteration is $M/2$ which is just
the midpoint of the allowed range $0$--$M$.  


\clearpage


\clearpage

\clearpage

\begin{deluxetable}{cccccccccccccc}
\rotate
\tabletypesize{\scriptsize}
\tablecaption{LOWER BOUND MASSES AND OTHER OUTPUT PARAMETERS FOR SN~2003fg
              FOR VARIOUS CHOICES OF INPUT PARAMETERS FOR THE SSC MODEL\label{table1}}
\tablewidth{0pt}   
\tablehead{
\colhead{Symbol\tablenotemark{a}}
&\colhead{$M_{\rm IPE\mathhyphen core}$\tablenotemark{b}}
&\colhead{$\rho_{\rm central,WD}$\tablenotemark{c}}
&\colhead{$g$\tablenotemark{d}}
&\colhead{$\alpha$\tablenotemark{e}}
&\colhead{$h$\tablenotemark{f}}
&\colhead{No. of\tablenotemark{g}}
&\colhead{$M_{\rm LB}$\tablenotemark{h}}  
&\colhead{$f_{\rm CO}$\tablenotemark{i}}  
&\colhead{$f_{\rm IME}$\tablenotemark{j}}  
&\colhead{$f_{\rm IPE}$\tablenotemark{k}}  
&\colhead{$f_{\rm ^{56}Ni}$\tablenotemark{l}}  
&\colhead{$v_{e}$\tablenotemark{m}}  
&\colhead{kinetic energy $E$\tablenotemark{n}}  \\
&\colhead{($M_{\odot}$)} 
&\colhead{($10^{9}\rm g\,cm^{-3}$)}
&&&
&\colhead{EPVs}
&\colhead{($M_{\odot}$)} 
&&&&
&\colhead{(${\rm km\,s^{-1}}$)} 
&\colhead{(foes)} \\
}
\startdata
  Y &1.84 & $\;\,$2.9         & 0.7  &1.2    & 1.0  & 0   & 2.84 & 0.000 & 0.350 & 0.650 & 0.455 & 2390 & 1.933 \\ 
  Y &1.84 & $\;\,$2.9         & 0.7  &1.2    & 0.0  & 1   & 2.44 & 0.245 & 0.000 & 0.755 & 0.528 & 2024 & 1.194 \\  
  X &1.30 & $\;\,$2.9         & 0.7  &1.7    & 1.0  & 1   & 2.17 & 0.000 & 0.402 & 0.598 & 0.419 & 2583 & 1.731 \\ 
  X &1.30 & $\;\,$2.9         & 0.7  &1.7    & 0.0  & 2   & 1.84 & 0.294 & 0.000 & 0.706 & 0.494 & 2191 & 1.055 \\  
  O &1.29 & $\;\,$2.9         & 1.0  &1.2    & 1.0  & 1   & 2.16 & 0.000 & 0.403 & 0.597 & 0.597 & 2587 & 1.726 \\  
  O &1.29 & $\;\,$2.9         & 1.0  &1.2    & 0.0  & 2   & 1.83 & 0.295 & 0.000 & 0.705 & 0.705 & 2195 & 1.052 \\ 
  N &0.91 & $\;\,$2.9         & 1.0  &1.7    & 1.0  & 2   & 1.68 & 0.000 & 0.457 & 0.543 & 0.543 & 2803 & 1.571 \\  
  N &0.91 & $\;\,$2.9         & 1.0  &1.7    & 0.0  & 3   & 1.40 & 0.348 & 0.000 & 0.652 & 0.652 & 2383 & 0.947 \\ 
  Y &1.84 &      10.0         & 0.7  &1.2    & 1.0  & 1   & 2.52 & 0.000 & 0.269 & 0.731 & 0.512 & 2104 & 1.332 \\  
  Y &1.84 &      10.0         & 0.7  &1.2    & 0.0  & 2   & 2.28 & 0.191 & 0.000 & 0.809 & 0.566 & 1840 & 0.920 \\  
  X &1.30 &      10.0         & 0.7  &1.7    & 1.0  & 2   & 2.02 & 0.000 & 0.354 & 0.646 & 0.452 & 2406 & 1.392 \\  
  X &1.30 &      10.0         & 0.7  &1.7    & 0.0  & 3   & 1.77 & 0.266 & 0.000 & 0.734 & 0.514 & 2095 & 0.929 \\  
  O &1.29 &      10.0         & 1.0  &1.2    & 1.0  & 2   & 2.00 & 0.000 & 0.356 & 0.644 & 0.644 & 2414 & 1.394 \\ 
  O &1.29 &      10.0         & 1.0  &1.2    & 0.0  & 3   & 1.76 & 0.268 & 0.000 & 0.732 & 0.732 & 2102 & 0.929 \\ 
  N &0.91 &      10.0         & 1.0  &1.7    & 1.0  & 3   & 1.62 & 0.000 & 0.438 & 0.562 & 0.562 & 2726 & 1.437 \\ 
  N &0.91 &      10.0         & 1.0  &1.7    & 0.0  & 4   & 1.414\tablenotemark{o} &0.356 & 0.000 & 0.644 & 0.644 & 2304 & 0.896\\ 
\enddata

\tablenotetext{a}{
Symbol is the symbol used for the data point on Figures~\ref{f1} and~\ref{f2}.
}

\tablenotetext{b}{
This is the IPE-core mass.
It depends inversely on the $g$ and $\alpha$ parameters.
In fact $M_{\rm IPE\mathhyphen core}g\alpha=1.55\,M_{\odot}$.
We make the approximation that the IPE core contains all explosion-synthesized IPEs
and that no explosion-synthesized IPEs are outside the IPE core.
}

\tablenotetext{c}{
The $\rho_{\rm central,WD}$ quantity is the central density of the progenitor CO WD.
}

\tablenotetext{d}{
The $g$ factor is the ratio of $^{56}$Ni mass to IPE mass which we also take to the IPE-core mass $M_{\rm IPE\mathhyphen core}$.
}

\tablenotetext{e}{
The $\alpha$ quantity is the bolometric-luminosity-maximum-light ratio of luminosity to the total instantaneous radioactive decay
energy release rate per unit $^{56}$Ni mass.
}

\tablenotetext{f}{
The $h=f_{\rm IME}/(f_{\rm CO}+f_{\rm IME})$ quantity is the fraction of non-IPE element matter in the ejecta that is
explosively-synthesized IME matter.
}

\tablenotetext{g}{
This is the number of extreme parameter values (EPVs) used in determining the lower bound mass.
}

\tablenotetext{h}{
The $M_{\rm LB}$ quantities are the lower bound masses for the given input parameters to the SSC~model.
}

\tablenotetext{i}{
The $f_{\rm CO}$ quantity is the mass fraction of CO in the ejecta. 
}

\tablenotetext{j}{
The $f_{\rm IME}$ quantity is the mass fraction of IMEs in the ejecta.  
}

\tablenotetext{k}{
The $f_{\rm IPE}$ quantity is the mass fraction of IPEs in the ejecta and also interior mass fraction of the
IPE core.  
}

\tablenotetext{l}{
The 
$f_{\rm ^{56}Ni}$
quantity is the mass fraction of $^{56}$Ni in the ejecta.
It equals $f_{\rm IPE}$ when $g=1$.
}

\tablenotetext{m}{
The $v_{e}$ quantity is the $e$-folding velocity of the ejecta.
}

\tablenotetext{n}{
This is the kinetic energy of the ejecta in foes ($1\,{\rm foe}=10^{51}\,{\rm ergs}$).
}

\tablenotetext{o}{
The last-line input parameters give no solution for the mass for $v_{\rm IPE\mathhyphen core}=8000\,{\rm km\,s^{-1}}$
since even a non-rotating WD for the
given input parameters is too massive to match this $v_{\rm IPE\mathhyphen core}$ value.
So we set $v_{\rm IPE\mathhyphen core}=7639.5\,{\rm km\,s^{-1}}$ which is the largest $v_{\rm IPE\mathhyphen core}$ for which
a solution exists.
The output mass is just the mass of a non-rotating WD.
It is an upper bound mass or close to an upper bound mass.
}

\tablecomments{
See the text \S~\ref{section-mass} for a full description the parameters (except $h$ and 
$f_{\rm ^{56}Ni}$), 
\S~\ref{section-SSC} for
a description of the SSC model and the $h$ parameter, and 
\S~\ref{section-comparison-sn2003fg} for a discussion of the parameter values and the table.
The SSC~model input parameters for the calculations of the lower bound masses are
IPE-core mass $M_{\rm IPE\mathhyphen core}$, IPE-core velocity ($8000\,{\rm km\,s^{-1}}$ for all cases, except for the
last line where $v_{\rm IPE\mathhyphen core}=7639.5\,{\rm km\,s^{-1}}$ was used),
$\rho_{\rm central,WD}$, and $h$.
The IPE-core mass was itself calculated from equation~(\ref{eq-IPE-core-mass}) (\S~\ref{section-mass})
using the $g$ and $h$ values
and H2006's value of $1.55\,M_{\odot}$ for 
$L_{\rm bol}/\dot E_{\rm ^{56}Ni}(t_{\rm bol})$.
The quantities $M_{\rm LB}$, $f_{\rm CO}$, $f_{\rm IME}$, $f_{\rm IPE}$, $f_{\rm ^{56}Ni}$, 
$v_{e}$, and $E$ are output parameters of the~SSC model. 
The output parameters were determined from the SSC model by the procedure given in Appendix~\ref{ap-SSC-mass}.
The IPE-core velocity $8000\,{\rm km\,s^{-1}}$ is an upper bound on the actual IPE-core velocity, and
so the output masses for this IPE-core velocity are lower bounds.
}
\end{deluxetable}


\begin{thebibliography}{}


\bibitem[Arnett(1979)]{arnett1979}
        Arnett, W. D. 1979, \apj, 230, L37 

\bibitem[Arnett(1982)]{arnett1982}
        Arnett, W. D. 1982, \apj, 253, 785 

\bibitem[Asplund et al.(2005)]{asplund2005}
         Asplund, M., Grevesse, N., \& Sauval, A. J. 2005, in ASP Conf. Ser. 336, Cosmic Abundances as Records of Stellar Evolution
         and Nucleosynthesis, ed. T. G. Barnes III \& F. N. Bash (San Francisco: ASP), 25, astro-ph/0410214

\bibitem[Baron et al.(2006)]{baron2006}
    Baron, E., Bongard, S., Branch, D., \&~Hauschildt, P. H. 2006, \apj, 645, 480

\bibitem[Baron \&~Cooperstein(1990)]{baron1990}
    Baron, E., \& Cooperstein, J.  1990, \apj, 353, 597 

\bibitem[Benetti et al.(2005)]{benetti2005} 
         Benetti, S., et al. 2005, \apj, 623, 1011 


\bibitem[Branch(1992)]{branch1992} Branch, D. 1992, \apj, 392, 35 

\bibitem[Branch et al.(2005)]{branch2005} Branch, D., Baron, E., Hall, N.,
         Melakayil, M., \&~Parrent, J. 2005, PASP, 117, 545

\bibitem[Branch et al.(2003)]{branch2003} Branch, D., et al. 2003, AJ, 126, 1489

\bibitem[Branch et al.(2006)]{branch2006} Branch, D., et al.
          2006, PASP, 118, 560

\bibitem[Central Bureau for Astronomical Telegrams(2006)]{cbat2006}
         Central Bureau for Astronomical Telegrams 2006, (Cambridge, Massachusetts:  Center for Astrophysics),
         http://cfa-www.harvard.edu/iau/cbat.html

\bibitem[Chandrasekhar(1935)]{chandrasekhar1935} Chandrasekhar, S. 1935, MNRAS, 95, 207 

\bibitem[Chandrasekhar(1957)]{chandrasekhar1957} Chandrasekhar, S. 1957,
        An Introduction to the Study of Stellar Structure (New York: Dover Publications, Inc.)

\bibitem[Clayton(1983)]{clayton1983} Clayton, D. D. 1983,
        Principles of Stellar Evolution and Nucleosynthesis (Chicago:  The University of Chicago Press)

\bibitem[Conley et al.(2006)]{conley2006}
       Conley, A., et al. 2006, \aj, 132, 1707 

\bibitem[Cox(2000)]{cox2000} Cox, A. N. (ed.) 2000,
        Allen's Astrophysical Quantities, 4th Edition (New York: AIP/Springer-Verlag)

\bibitem[Dwarkadas \&~Chevalier(1998)]{dwarkadas1998}
        Dwarkadas, V. V., \&~Chevalier, R. A. 1998, \apj, 497, 807 

\bibitem[Firestone \&~Ekstr\"om(2004)]{firestone} Firestone, R. B., \& Ekstr\"om, L. P. 2004,
        LBNL Isotopes Project-Lunds Universitet:  WWW Table of Radioactive Isotopes
        (Berkeley, California:  Lawrence Berkeley National Laboratory),
        http://ie.lbl.gov/toi/

\bibitem[Fisher et al.(1999)]{fisher1999} Fisher, A., Branch, D., Hatano, K., \&~Baron, E. 1999, 
       \mnras, 304, 67 

\bibitem[Gibson \&~Stetson(2001)]{gibson2001}
       Gibson, B. K., \&~Stetson, P. B.  2001, ApJ, 547, L103 

\bibitem[H{\"o}flich \& Khokhlov(1996)]{hoeflich1996} H\"oflich, P.,
         \&~Khokhlov, A. 1996, \apj, 457, 500   

\bibitem[H{\"o}flich et al.(1998)]{hoeflich1998} H\"oflich, P.,
         Wheeler, J. C., \&~Thielemann, F.-K. 1998, \apj, 495, 617  

\bibitem[Howell et al.(2001)]{howell2001} Howell, D. A., H\"oflich, P.,
    Wang, L., \&~Wheeler, J. C. 2001, \apj, 556, 302 

\bibitem[Howell et al.(2006)]{howell2006} Howell, D. A., et al. 2006,
    Nature, 443, 308 (H2006)

\bibitem[Jeffery(1999)]{jeffery1999} Jeffery, D. J. 1999, astro-ph/9907015

\bibitem[Jeffery et al.(1992)]{jeffery1992}
        Jeffery, D. J., Leibundgut, B., Kirshner, R. P., Benetti, S.,
        Branch, D., \&~Sonneborn, G. 1992, \apj, 397, 304

\bibitem[Khokhlov et al.(1993)]{khokhlov1993}
             Khokhlov, A., M\"uller, E., \& H\"oflich, P. 1993, A\&A, 270, 223

\bibitem[Kirshner et al.(1993)]{kirshner1993}
             Kirshner, R. P., et al. 1993, ApJ, 415, 589

\bibitem[Langer et al.(2000)]{langer2000}
    Langer, N., Deutschmann, A., Wellstein, S., \& H\"oflich, P. 2000, A\&A, 362, 1046

\bibitem[Livne \&~Glasner(1990)]{livne1990} Livne, E., \&~Glasner, A. 1990, \apj, 361, 244 

\bibitem[Livne \&~Glasner(1991)]{livne1991} Livne, E., \&~Glasner, A. 1991, \apj, 370, 272 


\bibitem[Mazzali et al.(1998)]{mazzali1998}
        Mazzali, P. A., Cappellaro, E., Danziger, I. J., Turatto, M., \&~Benetti, S. 1998, \apj, 499, L49 

\bibitem[Mazzali et al.(1997)]{mazzali1997}
        Mazzali, P. A., Chugai, N., Turatto, M., Lucy, L. B., Danziger, I. J., Cappellaro, E.,
        Della Valle, M., \&~Benetti, S. 1997, \mnras, 284, 151 


\bibitem[NIST(2005)]{nist2005} NIST (National Institute of Standards and Technology) 2005,
        Atomic Weights and Isotopic Compositions with Relative Atomic Masses  (Gaithersberg, Maryland:  NIST),
        http://physics.nist.gov/PhysRefData/Compositions/index.html

\bibitem[NIST(2006)]{nist2006a} NIST (National Institute of Standards and Technology) 2006,
        The NIST Reference on Constants, Units, and Uncertainties (Gaithersberg, Maryland:  NIST),
        http://physics.nist.gov/cuu/Constants/

\bibitem[Nomoto \& Kondo(1991)]{nomoto1991}
         Nomoto, K., \& Kondo, Y. 1991, \apj, 367, L19 

\bibitem[Nomoto et al.(1984)]{nomoto1984}
         Nomoto, K., Thielemann, F.-K., \&~Yokoi, K. 1984, \apj, 286, 644

\bibitem[Nugent et al.(1997)]{nugent1997}
         Nugent, P., Baron, E., Branch, D., Fisher, A., \&~Hauschildt, P. 1997 \apj, 485, 812

\bibitem[Ostriker \&~Bodenheimer(1968)]{ostriker1968} Ostriker, J. P, \&~Bodenheimer, P. 1968, \apj, 151, 1089 

\bibitem[Phillips(1993)]{phillips1993} Phillips, M. M. 1993, ApJ, 413, L105 

\bibitem[Phillips et al.(1999)]{phillips1999} Phillips, M. M. 
        Lira, P., Suntzeff, N. B., Schommer, R. A., Hamuy, M., \& Maza, J. 
        1999, AJ, 118, 1766 

\bibitem[Piersanti et al.(2003)]{piersanti2003}
        Piersanti, L., Gagliardi, S., Iben, I., Jr., \& Tornamb\`e, A. 2003, \apj, 598, 1229

\bibitem[Pizzochero(1990)]{pizzochero1990}
        Pizzochero, P. 1990, \apj, 354, 333 

\bibitem[Press et al.(1992)]{press1992} Press, W. H., Teukolsky, S. A.,
        Vetterling, W. T., \& Flannery, B. P. 1992,
        Numerical Recipes in Fortran
        (Cambridge:  Cambridge University Press),
         http://library.lanl.gov/numerical/index.html

\bibitem[Richardson et al.(2002)]{richardson2002}
    Richardson, D., Branch, D., Casebeer, D., Millard, J., Thomas, R. C., \&~Baron, E.  2002, \aj, 123, 745

\bibitem[Ruiz-Lapuente et al.(1995)]{ruiz-lapuente1995}
        Ruiz-Lapuente, P., Kirshner, R. P., Phillips, M. M., Challis, P. M., Schmidt, B. P., Filippenko, A. V., \&~Wheeler, J. C.
        1995, \apj, 439, 60

\bibitem[Saha et al.(2001)]{saha2001}
        Saha, A., Sandage, A., Thim, F., Labhardt, L., Tammann, G. A., Christensen, J., Panagia, N., \&~Macchetto, F. D.
        2001, \apj, 551, 973

\bibitem[Saio \&~Nomoto(2004)]{saio2004}
        Saio, H., \& Nomoto, K. 2004, \apj, 615, 444

\bibitem[Shapiro \&~Teukolsky(1983)]{shapiro1983} Shapiro, S. L., \& Teukolsky, S. A. 1983,
        Black Holes, White Dwarfs, and Neutron Stars:  The Physics of Compact Objects 
        (New York: John Wiley \& Sons)

\bibitem[Thielemann et al.(1986)]{thielemann1986} Thielemann, F.-K., Nomoto, K.,
        \&~Yokoi, K. A\&A, 158, 17

\bibitem[Tornamb\`e \&~Piersanti(2005)]{tornambe2005}
         Tornamb\`e, A., \& Piersanti, L. 2005, in ASP Conf. Ser. 342,  1604--2004:  Supernovae as Cosmological
         Lighthouses, ed. M.~Turatto, S.~Benetti, L.~Zamperi, \& W.~Shea (San Francisco: ASP), 169 

\bibitem[Wang et al.(2003)]{wang2003}
    Wang, L., et al. 2003, \apj, 591, 1110

\bibitem[Wiese et al.(1966)]{wiese1966}
            Wiese, W. L., Smith, M. W., \&~Glennon, B. M. 1966, in
           {NSRDS--NBS}~{4}, {Atomic Transition Probabilities:
           Vol.~I---Hydrogen through Neon} (Washington, DC:  Government Printing Office)

\bibitem[Wiese et al.(1969)]{wiese1969}
           Wiese, W. L., Smith, M. W., \&~Miles, B. M. 1969, in
           {NSRDS--NBS}~{22}, {Atomic Transition Probabilities:
           Vol.~2---Sodium through Calcium} (Washington, DC:  Government Printing Office)

\bibitem[Woosley(1991)]{woosley1991}
    Woosley, S. E. 1991, in Gamma-Ray Line Astrophysics,
    ed.~P.~Durouchoux \&~N.~Prantzos
    (Paris:  American Institute of Physics), 270 


\bibitem[Woosley et al.(2006)]{woosley2006}
    Woosley, S. E., Kasen, D., Blinnikov, S., \& Sorokina, E. 2006, \apj, submitted, astro-ph/0609562 

\bibitem[Woosley \&~Weaver(1994)]{woosley1994}
    Woosley, S. E., \&~Weaver, T. A. 1994, in Supernovae:
    Session~LIV of the Les~Houches \'Ecole d'\'Et\'e de Physique
    Th\'eorique, ed.~S.~A.~Bludman, R.~Mochkovitch, \&~J.~Zinn-Justin
    (Amsterdam:  North-Holland), 63

\bibitem[Yoon(2006)]{yoon2006}
    Yoon, S.-C. 2006, private communication 

\bibitem[Yoon \&~Langer(2005)]{yoon2005}
    Yoon, S.-C., \&~Langer, N. 2005, A\&A, 435, 967 (YL)
         
\end{thebibliography}
\end{document}